\documentclass[a4paper,10pt]{article}
\usepackage{graphicx}
\usepackage{pict2e}
\usepackage{epsfig}
\usepackage{amssymb}
\usepackage{amsmath}
\usepackage{amsfonts, latexsym,cite}
\usepackage{graphicx}
\usepackage[all]{xy}
\usepackage{multirow}
\begin{document}

\newcommand{\ds}{\displaystyle}
\newcommand{\be}{\begin{equation}}
\newcommand{\ee}{\end{equation}}
\newcommand{\ba}{\begin{array}}
\newcommand{\ea}{\end{array}}
\newcommand{\bea}{\begin{eqnarray}}
\newcommand{\eea}{\end{eqnarray}}
\newtheorem{example}{Example}
\newtheorem{theorem}{Theorem}
\def\R{{\mathbb R}}
\def\C{{\mathbb C}}


%
%

\title{Quadratic algebra contractions and 2nd order superintegrable systems }

\author{Ernest G. Kalnins\\
Department of Mathematics,
University
of Waikato,\\ Hamilton, New Zealand,\\
 math0236@waikato.ac.nz \\
W.\ Miller, Jr.\\
School of Mathematics, University of Minnesota,\\
 Minneapolis, Minnesota,
55455, U.S.A.\\
 miller@ima.umn.edu}

\maketitle


\begin{abstract} Quadratic algebras are generalizations of Lie algebras; they include the symmetry algebras of 2nd order superintegrable systems in 2 dimensions as special cases. The superintegrable  systems are exactly solvable physical systems in classical and quantum mechanics. For constant curvature spaces we show that the free quadratic algebras generated by the 1st and 2nd order elements in the enveloping algebras of their Euclidean and orthogonal symmetry algebras  correspond one-to-one with the possible superintegrable systems with potential defined on these spaces.  We describe a contraction theory for quadratic algebras and show that for constant curvature superintegrable systems, ordinary Lie algebra contractions  induce contractions of the quadratic algebras of the superintegrable systems that correspond to geometrical pointwise limits of the physical systems.
One consequence is that by contracting function space realizations of  representations of the generic superintegrable quantum system on the 2-sphere (which give the structure equations for  Racah/Wilson polynomials) to  the other  superintegrable systems one  obtains the 
full Askey scheme of orthogonal hypergeometric polynomials. 
\end{abstract}

Keywords: contractions; quadratic algebras; superintegrable systems; Askey scheme

Mathematics Subject Classification 2000: 22E70, 16G99, 37J35, 37K10, 33C45, 17B60

\section{Introduction} 
In this special issue honoring Frank Olver, a paper devoted to algebraic issues for superintegrable systems might seem out of place. 
However, there are very close connections with Frank's interests. Quantum superintegrable systems are explicitly solvable problems with physical interest
and special functions arise through this association. Most  special functions of
mathematical physics, as listed in the Digital Library of Mathematical Functions, appear via separation of variables, determined by 2nd order 
symmetry operators of superintegrable systems. 
Most  special functions that are solutions of 2nd order difference equations arise as function space realizations of 
representations of symmetry algebras of superintegrable systems. Orthogonal polynomials,  continuous and discrete, appear naturally
in this context.The structure theory of superintegrable systems 
provides a way of organizing special functions and relating their properties, an alternative approach to the DLMF.

For example, consider the following quantum superintegrable system:  the generic 3-parameter potential 
on the 2-sphere \cite{KKMP}. The eigenvalue equation $H\Psi=E\Psi$ separates in spherical coordinates (in multiple ways) and in 
Lam\'e coordinates \cite{KMJP}. The spherical coordinate  eigenfunctions are the orthogonal Prorial/Karlin-McGregor polynomials, orthogonal on a 
right triangle  \cite{KMT}. The corresponding eigenfunctions in  1-parameter function space realizations of the  irreducible representations of 
the symmetry algebra are the Racah and Wilson polynomials, in full generality \cite{KMPost11}.  If we set two of the parameters in the potential equal to 0 so that the 
restricted system has axial symmetry,
then the quantum system (the Higgs oscillator)  still separates in two spherical coordinate systems. 
One set of eigenfunctions yields the  
Koschmieder polynomials, the other the Zerneke polynomials, orthogonal on the unit disk \cite{KMT}. 
Corresponding function space realizations of the irreducible representations of the symmetry algebra yield Hahn and dual 
Hahn polynomials. Taking pointwise limits of this generic system we can contract it to a variety of quantum systems on flat space, with separable eigenfunctions expressed as products of Hermite, Laguerre and Jacobi polynomials for bound
 states, and
with continuous spectra associated with hypergeometric, confluent hypergeometric and Bessel functions. 
Taking  contractions of the irreducible 
function space realizations of the symmetry algebras and ``saving a representation'' in the sense of Wigner \cite{Wigner}, we can recover the Askey scheme 
for hypergeometric orthogonal polynomials \cite{KMPost13} and limit relations for more complicated functions, such as Lam\'e, Heun  and Mathieu functions \cite{KMP}.

Given an $n$-dimensional Riemannian or pseudo-Riemanian manifold, real or complex,  we define a  quantum Hamiltonian in local coordinates $x_i$ as 
$ H=\Delta_n+V({\bf x})\equiv \frac{1}{\sqrt{g}}\sum_{jk=1}^n\partial_j(g^{jk}\sqrt{g}\partial_k) +V({\bf x})$ where $\Delta_n$ is the
Laplace-Beltrami operator in these coordinates,  $g^{jk}({\bf x})$ is the contravariant metric tensor and $g$ is the determinant 
of the covariant metric tensor. $V$ is a scalar potential.
  The  quantum system is (maximally) {\it superintegrable} if there are 
    $2n-1$ 
algebraically independent  partial differential operators $L_1,\cdots,L_{2n-2},L_{2n-1}=H$ that commute with $H$. All functions of the coordinates are assumed 
locally analytic.
Similarly a classical Hamiltonian 
${\cal H} =\sum _{jk}g^{jk}p_jp_k+V(\bf x)$ is {\it superintegrable} if there are 
    $2n-1$ 
functionally independent constants of the motion  ${\cal L}_1,\cdots,{\cal L}_{2n-2},{\cal L}_{2n-1}={\cal H}$ in involution with  
$\cal H$: $\{{\cal L}_\ell,{\cal H}\}=0$, $\ell=1,\cdots,2n-1$, with respect to the Poisson bracket $ \{{\cal F}({\bf p},{\bf x}),{\cal G}({\bf p},{\bf x})\}=\sum_{j=1}^n\left( \partial_{p_j} {\cal F}\ \partial_{x_j}{\cal G}
-\partial_{p_j} {\cal G}\ \partial_{x_j}{\cal F}\right)$. (Throughout the paper we use $\cal L$ for constants of the motion and $L$ for quantum symmetries.)

It is assumed that the ${\cal L}_\ell$ are polynomial 
functions of the momenta $p_j$ and globally defined in the $x_j$ except for possible singularities on lower dimensional manifolds.
The maximum possible number of functionally independent constants of the motion is $2n-1$ and this 
maximum is rarely achieved. Superintegrability captures the properties of 
quantum Hamiltonian systems that allow the Schr\"odinger eigenvalue problem $H\Psi=E\Psi$ to be solved exactly, analytically and 
algebraically and the orbits of the classical superintegrable systems to be determined algebraically. For a more careful discussion of superintegrability and its applications, see \cite{MPW}. 

The key to the connection between solvability and superintegrability lies in the symmetry algebra $ S$ produced from 
the generators $L_\ell$ by taking linear combinations, products and commutators.  If a system is merely integrable with $n$ {\it commuting} generators $L_j$
then the  algebra is abelian. However it is not possible to have more than $n$ commuting 
independent operators, so for a superintegrable system the symmetry algebra is necessarily nonabelian.  Since $ S$ maps each energy eigenspace of $H$ into 
itself  the eigenspaces are multiply degenerate, and the irreducible representations of $S$ give  the
possible degeneracies and  energy eigenvalues.

 A quantum  system is  of order $K$ if the maximum order of the symmetry
 operators, other than $H$, is  $K$. (There is a similar definition for classical systems, based on the order of the symmetries as polynomials in the momenta.)   Much of the recent excitement in superintegrability theory is
due to the discovery of superintegrable systems for  $n$ and $K$ arbitrarily large, e.g., \cite{TTW, TTW2, PW2011}, with no  connection between these systems and group theory. 
However, for  $n=2$, $K=1,2$   a connection exists.

In \cite{KMPost13} the concept of a contraction of the symmetry algebra of a 2D 2nd order superintegrable system was introduced  and the Askey scheme 
as  derived via contractions. However,  it was unclear how the contractions were found;  the procedure 
appeared complicated. Here we demonstrate that all of the limits are induced by Wigner-In\"on\"u  contractions of
the Lie algebras $e(2,\C)$ and $o(3,\C)$, already classified. Further, all of the quadratic algebras of 2nd order 2D superintegrable systems 
correspond 1-1 to free quadratic algebras contained in the enveloping algebras of  $e(2,\C)$ and $o(3,\C)$. Thus, though many of these  systems admit 
no group symmetry, their structures are determined by the underlying Lie algebras.

\section{2D 2nd order superintegrability}
For $n=2$, $K=2$,  a superintegrable system admits 3 symmetries and in this special case there is a $1-1$ relation between quantum and classical symmetries, \cite{KKM20061}. The potentials are the same and corresponding to a 2nd order classical constant of the motion  ${\cal L}=\sum_{j,k=1}^2L^{jk}({\bf x})p_jp_k+W({\bf x})$,  $L^{jk}$ a symmetric contravariant tensor, the quantum symmetry is $L=\frac{1}{\sqrt{g}}\sum_{ij=1}^2\partial_i(L^{ij}({\bf x})\sqrt{g}\partial_j)+W({\bf x})$. Here $L$ is formally self-adjoint with respect to the bilinear form 
$ <f_1,f_2>_g=\int f_1({\bf x})f_2({\bf x})\sqrt{g({\bf x})}\ dx_1 dx_2
$
on the manifold,\cite{KMPost1}.
The set $\{H,L_1,L_2\}$ of generating symmetries is required to be algebraically independent, i.e., there is no nontrivial
polynomial $P(H,L_1,L_2)$, symmetric in $L_1,L_2$ such that $P\equiv 0$. For our treatment of 2nd order
2D quantum  systems
the values of the mass $m$ and Planck's
constant $\hbar$ are immaterial, so we have normalized our Hamiltonians as given.
Every 2D Riemannian space is conformally flat so there exist  Cartesian-like coordinates
$x_1, x_2$ such that
\be\label{quantumsupint} H=\frac{1}{\lambda({\bf x})}(\partial_{11}+\partial_{22})+V({\bf x}),\
L_\ell=\frac{1}{\lambda}\sum_{j,k=1}^2\partial_{j}\left(L^{jk}_{(\ell)}\lambda)\partial_k\right)+W_{(\ell)}
({\bf x}),\ k=1,2,\ \ee 
(A 1st order constant of the motion  ${\cal X}=\sum_{j=1}^2f_j({\bf x})p_j$ corresponds to the  formally skew-adjoint  symmetry operator $X=\sum_{j=1}^2\left(f_j\partial_{x_j}+\frac{\partial_{x_j}(\lambda f_j)}{2\lambda}\right)$.)
The symmetry  relations $\{{\cal H},{\cal L}_k\}=0$, $k=1,2$, put conditions on the functions $W_{(1)},W_{(2)}$. If we require that the symmetries are linearly functionally independent, i.e., that $g_1L_1+g_2L_2+g_3H\equiv 0$  for functions $g_j$ implies $g_1\equiv g_2\equiv g_3 \equiv 0$ , we can 
solve for the partial derivatives $\partial_jW_{(k)}$
in terms of the function $V$ and its 1st derivatives. The integrability conditions
$\partial_1(\partial_2W_{(k)})=\partial_2(\partial_1W_{(k)})$,
the Bertrand-Darboux equations \cite{KKM20041}, lead to the necessary and sufficient condition that $V$ must satisfy a
pair of coupled linear equations of the form
\be\label{canonicalequations} V_{22}-V_{11}=A^{22}V_1+B^{22}V_2,\quad V_{12}=A^{12}V_1+B^{12}V_2,\ee
for locally analytic functions $A^{ij}({\bf x}),B^{ij}({\bf x})$. Here $V_i=\partial_iV$, etc.
We call these the {\it canonical equations}. If the integrability equations for
(\ref{canonicalequations}) are satisfied identically then the solution space  is 4-dimensional
and we can always express the
 solution in the form $V({\bf x})= \sum_{j=1}^3a_jV_{(j)}({\bf x})+a_4$
where $a_4$ is a trivial additive constant. In this case  the potential is {\it nondegenerate} and 
3-parameter.
Another possibility is that the solution space is 2-dimensional
with general solution
$V({\bf x})= a_1V_{(1)}({\bf x})+a_2$.
 Then  the potential is {\it degenerate} and  1-parameter.
Every degenerate potential can be obtained from some nondegenerate
 potential by parameter restriction, \cite{fine}. It is not {\it just} a restriction, however, because the
symmetry algebra changes. A  formally skew-adjoint 1st order symmetry appears and this induces a new 2nd order symmetry.
A third possibility is that the integrability conditions are satisfied only by a constant potential.
In that case we refer to the system as {\it free}; the free equation $H\Psi=E\Psi$ is just the Laplace-Beltrami eigenvalue equation.
Note: Any 2-parameter potential extends to a 3-parameter potential, \cite{fine}. There is one remaining 
possibility: we can satisfy relations  $[ H,L_k]=0$, but the symmetries $L_1,L_2$ are functionally linearly dependent. There is a single exceptional superintegrable system for which this is true, $E_{15}$ in our listing \cite{KKMP}. 
All of the systems with nondegenerate  potential (and $E_{15}$)  have the remarkable property that the symmetry algebras generated by $H,L_1,L_2$ 
 close polynomially under commutation, as follows.
Define the 3rd order commutator $R$ by $R=[L_1,L_2]$. Then the fourth order operators $[R,L_1],[R,L_2]$
are contained in the associative algebra of
symmetrized products of the generators \cite{KKM20041}:\hfill\break
\be\label{commutator}[L_j,R]=\sum_{0\leq e_1+e_2+e_3\leq 2} M^{(j)}_{e_1,e_2,e_3} \{ L_1^{e_1}, L_2^{e_2}\} H^{e_3},
 \qquad e_k\geq 0,\ L_k^0=I,\ee
where  $\{L_1,L_2\}=L_1L_2+L_2L_1$ is the symmetrizer.
Also the 6th order operator $R^2$ is contained in the algebra of symmetrized products up to 3rd order:
\be\label{Casimir1} R^2 -\sum_{0\leq e_1+e_2+e_3\leq 3} N_{e_1,e_2,e_3} \{ L_1^{e_1}, L_2^{e_2}\} H^{e_3} = 0.\ee
In both equations the constants $M^{(j)}_{e_1, e_2, e_3}$ and $N_{e_1, e_2, e_3}$ are polynomials in the parameters
$a_1, a_2, a_3$ of degree $2-e_1-e_2-e_3$ and $3-e_1-e_2-e_3$, respectively.

For systems with one parameter potentials, \cite{fine},
 there are 4 generators: one 1st order ${\cal X}$ and three 2nd order $H,L_1,L_2$.
The commutators $[X,L_1],[ X,L_2]$ are 2nd order and expressed as
 \be\label{structure2}[X,L_j]=\sum_{0\leq e_1+e_2+e_3+e_4\leq 1} P^{(j)}_{e_1,e_2,e_3, e_4} L_1^{e_1}L_2^{e_2}H^{e_3}X^{e_4} ,\quad j=1,2.\ee
The commutator $[L_1,L_2]$ is 3rd order, skew adjoint, and expressed as
\be\label{commutator1}[L_1,L_2]=\sum_{0\leq e_1+e_2+e_3+e_4\leq 1} Q_{e_1,e_2, e_3, e_4}\{ L_1^{e_1}L_2^{e_2},X\}H^{e_3}X^{2e_4}. \ee
Finally,  there is a 4th order relation:
\be\label{Casimir2}G\equiv\sum_{0\leq e_1+e_2+e_3+e_4\leq 2} S_{e_1, e_2, e_3, e_4}\{L_1^{e_1},L_2^{e_2}, X^{2e_4}\}H^{e_3}=0, \ X^0=H^0=I,\ee
where $\{L_1^{e_1},L_2^{e_2}, X^{2e_4}\}$ is the 6-term symmetrizer of three operators. The constants
$P^{(j)}_{e_1, e_2, e_3, e_4}$, $Q_{e_1, e_2, e_3, e_4}$ and $S_{e_1, e_2, e_3, e_4}$
are polynomials in the parameter $a_1$ of degrees  $1-e_1-e_2-e_3-e_4,$  $1-e_1-e_2-e_3-e_4$  and $2-e_1-e_2-e_3-e_4$, respectively.

We call these symmetry algebras for degenerate and nondegenerate systems {\it quadratic algebras}, in the sense that the commutators of the
generators are at most
quadratic expansions in the generators. Usually, the generators for free systems form an algebra that doesn't close,  not a quadratic algebra. 

There is an analogous quadratic algebra structure for classical superintegrable systems in 2D.
All these classical systems have the  property that the symmetry algebras generated by ${\cal H},{\cal L}_1,{\cal L}_2$ for
nondegenerate potentials close under Poisson brackets.
Define the 3rd order bracket $\cal R$ by ${\cal R}=\{{\cal L}_1,{\cal L}_2\}$. Then the fourth order constants of the motion
$\{{\cal L}_j,{\cal R}\}$
are can be expressed as, \cite{KKM20041}:\hfill\break
\be\label{Bracket}\{{\cal L}_j,{\cal R}\}=\sum_{0\leq e_1+e_2+e_3\leq 2} M^{(j)}_{e_1,e_2,e_3} {\cal  L}_1^{e_1}{\cal L}_2^{e_2}{\cal  H}^{e_3},
 \qquad e_k\geq 0,\ {\cal L}_k^0=1.\ee
Also the 6th order constant of the motion ${\cal R}^2$  satisfies:
\be\label{Casimirclassr2} {\cal R}^2 -\sum_{0\leq e_1+e_2+e_3\leq 3} N_{e_1,e_2,e_3} {\cal  L}_1^{e_1}{\cal  L}_2^{e_2}{\cal  H}^{e_3} = 0.\ee
In both equations the constants $M^{(j)}_{e_1, e_2, e_3}$ and $N_{e_1, e_2, e_3}$ are polynomials in the parameters
$a_1, a_2, a_3$ of degree $2-e_1-e_2-e_3$ and $3-e_1-e_2-e_3$, respectively.

For one parameter potentials, \cite{fine},
 there are 4 generators: one 1st order in momenta  $\cal X$ and three 2nd order ${\cal H},{\cal L}_1, {\cal L}_2$.
The brackets $\{{\cal X},{\cal L}_j\}$ are 2nd order:
 \be\label{structure2p}\{{\cal X},{\cal L}_j\}=\sum_{0\leq e_1+e_2+e_3+e_4\leq 1} P^{(j)}_{e_1,e_2,e_3, e_4}
{\cal L}_1^{e_1}{\cal L}_2^{e_2}{\cal H}^{e_3}{\cal X}^{e_4} ,\quad j=1,2.\ee
The bracket $\{{\cal L}_1,{\cal L}_2\}$ is 3rd order  and expressed as
\be\label{Bracket1p}\{{\cal L}_1,{\cal L}_2\}=\sum_{0\leq e_1+e_2+e_3+e_4\leq 1} Q_{e_1,e_2, e_3, e_4}
{\cal  L}_1^{e_1}{\cal L}_2^{e_2}{\cal X}{\cal H}^{e_3}{\cal X}^{2e_4}. \ee
There  is  a 4th order relation obeyed by the generators:
\be\label{Bracket2p}{\cal G}\equiv\sum_{0\leq e_1+e_2+e_3+e_4\leq 2} S_{e_1, e_2, e_3, e_4}{\cal L}_1^{e_1}{\cal L}_2^{e_2}{\cal  X}^{2e_4}{\cal H}^{e_3}=0,
\ {\cal X}^0={\cal H}^0=1.\ee The constants
$P^{(j)}_{e_1, e_2, e_3, e_4}$, $Q_{e_1, e_2, e_3, e_4}$ and $S_{e_1, e_2, e_3, e_4}$
are polynomials in  $a_1$ of degrees  $1-e_1-e_2-e_3-e_4,$  $1-e_1-e_2-e_3-e_4$  and $2-e_1-e_2-e_3-e_4$, respectively.

For free systems that do not admit  a 1- or 3-parameter  potential  the algebra of the generators
normally doesn't close, see  \S \ref{section4.1}. The structure equations for the quadratic algebras of associated classical and quantum systems are not identical, but they agree in the highest order terms.
The differences are  1) quantum operators may not commute and for quantization,
products of constants of the motion  are replaced by operator symmetrizers,
 and 2) even order symmetry operators in the generating basis must be formally self-adjoint; odd order ones formally skew-adjoint.

We can study quadratic algebras in general, whether or not they arise as symmetry algebras of a superintegrable system. Thus, we define an abstract {\it nondegenerate (quantum) quadratic algebra}
is a noncommutative associative  algebra generated by linearly independent operators $H, L_1,L_2$, with parameters $a_1,a_2,a_3$,
such that $H$ is in the center and relations (\ref{commutator}), (\ref{Casimir1}) hold. Similarly we define an abstract {\it degenerate (quantum) quadratic algebra} 
is a noncommutative  associative  algebra generated by linearly independent operators $X, H, L_1,L_2$, with parameter $a_1$,
such that $H$ is in the center and relations (\ref{structure2}),(\ref{commutator1}),(\ref{Casimir2}) hold. We also consider systems  where all of the parameters $a_j$ are identically zero; these are {\it free nondegenerate} and {\it free degenerate} (quantum) quadratic algebras.  Analogously, 
an abstract {\it nondegenerate (classical) quadratic algebra}
is a  Poisson  algebra with functionally independent generators  ${\cal H}, {\cal L}_1,{\cal L}_2$, and parameters $a_1,a_2,a_3$,
such that all generators are in involution with {\cal H} and relations (\ref{Bracket}) and (\ref{Casimirclassr2}) hold. An abstract {\it degenerate (classical) quadratic algebra} 
is a Poisson algebra with linearly independent generators  ${\cal X}, {\cal H}, {\cal L}_1,{\cal L}_2$, and parameter $a_1$,
such that all generators are in involution with {\cal H} and relations (\ref{structure2p}),(\ref{Bracket1p}) and (\ref{Bracket2p}) hold. Systems  with all $a_j$ identically zero are {\it free nondegenerate} and {\it free degenerate} (classical) quadratic algebras.

\subsection{Nondegenerate classical structure equations}
Suppose the 2D classical second order superintegrable system with nondegenerate potential has 2nd order generators ${\cal L}_1,{\cal L}_2,{\cal H}$
with ${\cal R}=\{{\cal L}_1,{\cal L}_2\}$. The Casimir is ${\cal R}^2-F({\cal L}_1,{\cal L}_2,{\cal H},a_1,a_2,a_3)=0$ where the $a_j$ are the parameters
in the potential. It is easy to show that 
$\{{\cal L}_1,{\cal R}\}=\frac12\frac{\partial F}{\partial {\cal L}_2},\quad \{{\cal L}_2,{\cal R}\}=-\frac12\frac{\partial F}{\partial {\cal L}_1}$, so the Casimir contains within itself all of the structure equations. A similar, but more complicated result for nondegenerate quantum quadratic algebras will appear in a forthcoming paper.

\subsection{Degenerate classical structure equations}\label{degstructure}
Now suppose the 2D classical second order superintegrable system with degenerate (1-parameter) potential has generators ${\cal X}$ (1st order),
 and ${\cal L}_1,{\cal L}_2,{\cal H}$ (2nd order) with   Casimir $ G( {\cal X},{\cal L}_1,{\cal L}_2,{\cal H},\alpha)=0$, where the $\alpha$ is 
the parameter in the potential. Note that $G$ is determined only to within a multiplicative constant.  Now
$0=\{{\cal X},G\}=\frac{\partial G}{\partial {\cal L}_1}\{{\cal X},{\cal L}_1\}+\frac{\partial G}{\partial {\cal L}_2}\{{\cal X},{\cal L}_2\}$
$ \Longrightarrow \{{\cal X},{\cal L}_1\}=K\frac{\partial G}{\partial {\cal L}_2},\quad  \{{\cal X},{\cal L}_2\}=-K\frac{\partial G}{\partial {\cal L}_1}$,
for some constant  $K$, since $\{{\cal X},{\cal L}_j\}$, ${\partial G}/\partial{\cal L}_j$ are all 2nd order in the momenta. Further
$0=\{{\cal L}_1,G\}=\frac{\partial G}{\partial {\cal X}}\{{\cal L}_1,{\cal X}\}+\frac{\partial G}{\partial {\cal L}_2}\{{\cal L}_1,{\cal L}_2\}$,
$0=\{{\cal L}_2,G\}=\frac{\partial G}{\partial {\cal X}}\{{\cal L}_2,{\cal X}\}+\frac{\partial G}{\partial {\cal L}_1}\{{\cal L}_2,{\cal L}_1\}$.
Assuming $G$ depends nontrivially on at least one of ${\cal L}_1$, ${\cal L}_2$, we have
\be\label{Geqn2}\{{\cal X},{\cal L}_1\}=K\frac{\partial G}{\partial {\cal L}_2},\  \{{\cal X},{\cal L}_2\}=-K\frac{\partial G}{\partial {\cal L}_1},\
 \{{\cal L}_1,{\cal L}_2\}=K\frac{\partial G}{\partial {\cal X}}.\ee
Thus the structure equations are determined by $G$ to within a  constant.

For a degenerate superintegrable system it would seem that  it is possible that $K$ is a rational constant of the motion; either 1) the ratio of two 2nd
order polynomials in the momenta (necessarily two 2nd order constants of the motion) or 2) the ratio of two 1st order polynomials
in the momenta (necessarily multiples of ${\cal X}$. However, in case 1) it is easy to see that  this would imply 3
mutually involutive symmetries, impossible for a 2D Hamiltonian
system and case 2) is trivially equivalent  to a constant $K$. Thus for a 2D degenerate  superintegrable system $K$ is always a nonzero constant.
However, for free superintegrable systems rational $K$  can  occur.
\begin{example}
 For some functions ${\cal X},{\cal L}_1,{\cal L}_2, {\cal H}$ satisfying a polynomial relation $G=0$, $K$ may be rational.
For example,  the flat space system
 ${\cal L}_1= {\cal J}p_1,\ {\cal L}_2=p_1^2,\ {\cal X}={\cal J},\ {\cal H}=p_1^2+p_2^2$,
with ${\cal J}=xp_2-yp_1$, gives $G=-{\cal L}_1^2+{\cal X}^2{\cal L}_2=0$,  $K=-p_2/{\cal X}$. However,
this is not a degenerate superintegrable system. It is free.  
\end{example}

Degenerate superintegrable systems  are restrictions of the 3-parameter potentials  to 1-parameter ones, such that new symmetries appear:
 We can take a particular basis of 2nd order generators
${\cal H}, {\cal L}_1,{\cal L}_2$, and parameters $a_1,a_2,a_3$ for the classical physical system with nondegenerate potential, such that for $a_2=a_3=0$
the symmetry ${\cal L}_1$ becomes a perfect square:
${\cal L}_1|_{a_1=a_2=0}={\cal X}^2$.
Then ${\cal X}$ will be a 1st order symmetry for ${\cal H}_0={\cal H}|_{a_1=a_2=0}$ with no potential term, i.e., a Killing vector.
Noting the relation
$ {\cal R}=\{{\cal L}_1,{\cal L}_2\}=2{\cal X}\{{\cal X},{\cal L}_2\}$ upon restriction,
we see that ${\cal L}_3\equiv \{{\cal X},{\cal L}_2\}$ is a 2nd order symmetry for ${\cal H}_0$
 (usually linearly independent of the symmetries we already know). We can factor $2{\cal X}^2$ from each
term of the restricted identity ${\cal R}^2-{\cal F}=0$ to obtain the Casimir ${\cal G}=0$ for the contracted system, where ${\cal G}={\cal L}_3^2+\cdots$.
 In the limit,   (\ref{Geqn2}) (with ${\cal L}_1$ replaced by $ {\cal L}_3$) holds with constant $K$.

If however,  ${\cal L}_3$ is a linear combination of ${\cal X}^2,{\cal L}_2,{\cal H}_0$ then the resulting expression is identically satisfied and we get no additional
information about the degenerate structure algebra. By inspection one can verify that all Casimirs $G=0$ can be obtained as limits of equations
${\cal R}^2-{\cal F}=0$ for some nondegenerate superintegrable system, except for degenerate systems St\"ackel equivalent to $E_4$ or $E_{13}$, see below.
For those systems the new 2nd order symmetries
appear in a discontinuous manner. All these results have quantum analogies, as we shall show in a forthcoming paper.

\section{Free 2D 2nd order  superintegrable systems}
As was shown in \cite{KKM20042,KKM20061}  the `free'' 2nd order superintegrable system obtained by setting all the parameters in a nondegenerate potential
equal to zero retains
all of the information needed to reconstruct the potential. Thus we can, in principle, restrict our
attention to free systems. Here we explore this concept in more detail and extend it.
First we review how the structure equations for 2D 2nd order nondegenerate classical superintegrable systems are
determined. Such a system  admits a symmetry ${\cal L} =\sum
a^{ij}p_ip_j+W$ if and only if the Killing equations are satisfied
\be \label{Killingeqns}a^{ii}_i= -\frac{\lambda_1}{\lambda}a^{i1} -\frac{\lambda_2}{\lambda}a^{i2},\  i=1,2,\quad
2a^{ij}_i+a^{ii}_j=-\frac{\lambda_1}{\lambda}a^{j1} -\frac{\lambda_2}{\lambda}a^{j2},\ i,j=1,2,\ i\ne j,
\ee
where $a^{ij}_k=\partial_{x_k}a^{ij}$, as well as
$W_i=\lambda\sum_{j=1}^2 a^{ij}V_j$. Here $W_i=\partial_{x_1}W$ with a similar convention for subscripts on $V$.
The equations for $W$ can be solved provided the Bertrand-Darboux equation  $\partial_{x_1}W_{2}=\partial_{x_2}W_1$ holds. We can
solve the two independent  Bertrand-Darboux equations for the potential to obtain
the canonical system (\ref{canonicalequations}) 
where the $A^{ij},B^{ij}$ are computable from the generating constants of the motion.
For nondegenerate superintegrability, the integrability conditions for the canonical equations 
must be satisfied identically, so that  $V, V_1,V_2,V_{11}$ can be prescribed arbitrarily at a fixed regular point.

To obtain the integrability conditions for  equations  (\ref{canonicalequations})  we introduce the dependent variables
$W^{(1)}=V_1$, $W^{(2)}=V_2$,
$W^{(3)}=V_{11}$,  and matrices
\be\label{wvector1}{\bf w}=\left(\ba{c} W^{(1)}\\ W^{(2)}\\W^{(3)}\ea\right),\
{\bf A}^{(1)}=\left(\ba{ccc} 0&0&1\\ A^{12}&B^{12}&0\\A^{13}&B^{13}&B^{12}-A^{22}\ea\right),\
{\bf A}^{(2)}=\left(\ba{ccc} A^{12}&B^{12}&0 \\A^{22}&B^{22}&1  \\A^{23}&B^{23}&A^{12}\ea\right),
\ee
\bea
 A^{13}&=&A^{12}_2-A^{22}_1+B^{12}A^{22}+A^{12}A^{12}-B^{22}A^{12}\nonumber\\
B^{13}&=&B^{12}_2-B^{22}_1+A^{12}B^{12},\
  A^{23}= A^{12}_1+B^{12}A^{12},\quad  B^{23}=B^{12}_1+B^{12}B^{12}.\nonumber
\eea
Then the   integrability conditions for  system
$
\partial_{x_j}{\bf w}={\bf A}^{(j )}{\bf w},\  j=1,2$,
must hold:
\be\label{2int3}
A^{(2)}_1-A^{(1)}_2=A^{(1)}A^{(2)}-A^{(2)}A^{(1)}\equiv [A^{(1)},A^{(2)}].
\ee
If and only if   (\ref{2int3}) holds, the system has a 4D vector space of solutions $V$.

{}From the conditions that ${\cal L}$ is a constant of the motion
and relations  (\ref{canonicalequations})   we can solve for
all of the first partial derivatives $\partial_i( a^{jk})$ to obtain
\bea \label{2Dsym}\partial_1a^{11}&=&-G_1a^{11}-G_2 a^{12},\quad \partial_2 a^{22}=-G_1a^{12}-G_2a^{22},\\
3\partial_2  a^{12}&=&-3G_2a^{12}+( a^{11}-a^{22})(-B^{12}-G_1)+a^{12}(-B^{22}+G_2),\nonumber\\
3\partial_1 a^{22}&=&-3G_1a^{22}+(a^{11}- a^{22})(2B^{12}-G_1)+a^{12}(2B^{22}+G_2),\nonumber\\
3\partial_1 a^{12}&=&-3G_1a^{12}+( a^{11}- a^{22})(A^{12}+G_2)+a^{12}(A^{22}+G_1),\nonumber\\
3\partial_2 a^{11}&=&-3G_2a^{11}+(a^{11}-a^{22})(-2A^{12}+G_2)+ a^{12}(-2A^{22}+G_1),\nonumber
\eea
where $\lambda=\exp G$.
This system closes, so the space of solutions is exactly 3 dimensional. 
Note that if ${\cal L}_1=\sum_{k,j=1}^2\ell^{kj}(x,y)p_kp_j+W_{(1)}(x,y)$, ${\cal L}_2=\sum_{k,j=1}^2b^{kj}(x,y)p_kp_j+W_{(2)}(x,y)$, ${\cal L}_3={\cal H}$, is
 a basis for the symmetries then
\be\label{canoneqns} A^{12}=-G_2+\frac{D_{(2)}}{D},\quad A^{22}=2G_1+\frac{D_{(3)}}{D},\quad B^{12}=-G_1-\frac{D_{(0)}}{D},\quad B^{22}=-2G_2-\frac{D_{(1)}}{D},\ee
\[ D=\det \left(\ba{cc} \ell^{11}-\ell^{22},& \ell^{12}\\ b^{11}-b^{22},& b^{12}\ea\right),\quad D_{(0)}=\det \left(\ba{cc} 3\ell^{12}_2,& -\ell^{12}\\ 3b^{12}_2,& -b^{12}
\ea\right), \]
\[ D_{(1)}=\det \left(\ba{cc} 3\ell^{12}_2,& \ell^{11}-\ell^{22}\\ 3b^{12}_2,& b^{11}-b^{22}\ea\right),\quad D_{(2)}=\det \left(\ba{cc} 3\ell^{12}_1,& \ell^{12}\\
3b^{12}_1,& b^{12}\ea\right), \
D_{(3)}=\det \left(\ba{cc} 3\ell^{12}_1,& \ell^{11}-\ell^{22}\\ 3b^{12}_1,& b^{11}-b^{22}\ea\right).\]
The functions $A^{22},B^{22},A^{12},B^{12}$ are defined independent of the choice of basis for the 2nd order symmetries.
To determine the integrability conditions for system (\ref{2Dsym}) we define the vector-valued function
${\bf h}^{\rm tr}(x,y,z)=\left( a^{11},a^{12},a^{22}\right)$
and directly compute the $3\times 3$ matrix functions ${\cal A}^{(j)}$ to get the first-order system
$\partial_{x_j}{\bf h}={\cal A}^{(j )}{\bf h},\ j=1,2$, 
the integrability conditions for which are
\be\label{2int5}
{\cal A}^{(2)}_1-{\cal A}^{(1)}_2={\cal A}^{(1)}{\cal A}^{(2)}-{\cal A}^{(2)}{\cal A}^{(1)}\equiv [{\cal A}^{(1)},{\cal A}^{(2)}],
\ee
satisfied identically for a nondegenerate superintegrable system.

There is a similar analysis for a ``free'' 2nd order superintegrable system obtained by setting  the parameter in a {\it degenerate} potential
equal to zero, \cite{fine}: The free system  retains
all of the information needed to reconstruct the potential. All such degenerate superintegrable systems with potential are restrictions of nondegenerate
systems obtained by restricting the parameters 
so that one 2nd order symmetry becomes a perfect square, e.g. ${\cal L}_1={\cal X}^2$. Then ${\cal X}$ is a 1st order constant,
necessarily of the form
${\cal X}=\xi_1p_1+\xi_2 p_2$, without a function term. Since the degenerate systems are obtained by restriction, 
the potential function must satisfy
the equations (\ref{canonicalequations}) inherited from the nondegenerate system, with the same functions $A^{ij},B^{ij}$. In addition the relation
$\{{\cal X},{\cal H}\}=0$ imposes the condition $\xi_1V_1+\xi_2V_2=0$. By relabeling the coordinates, we can always assume $\xi_2\ne 0$ and
 write the system of equations for the
potential in the form
$V_2 = C^2V_1,\  V_{22}=V_{11} + C^{22}V_1,\  V_{12} = C^{12}V_1$,
where
 \[ C^2(x_1,x_2)=-\frac{\xi_1}{\xi_2},\ C^{22}(x_1,x_2)=A^{22} -\frac{\xi_1}{\xi_2}B^{22},\ C^{12}(x_1,x_2)= A^{12}-\frac{\xi_1}{\xi_2}B^{12}.\]
To find integrability conditions for these equations we introduce  matrices
\be\label{wvector3} {\bf v}=\left(\ba{c} V\\ V_1\ea\right),\
{\bf B}^{(1)}=\left(\ba{cc} 0&1\\ 0&\partial_2 C^2+C^2C^{12}-C^{22}\ea\right),\
{\bf B}^{(2)}=\left(\ba{cc} 0&C^2\\ 0&C^{12}\ea\right).
\ee
Then   integrability conditions for  system $\partial_{x_j}{\bf v}={\bf B}^{(j )}{\bf v}$, $j=1,2$,
must hold:
\be\label{2int9}
B^{(2)}_1-B^{(1)}_2=B^{(1)}B^{(2)}-B^{(2)}B^{(1)}\equiv [B^{(1)},B^{(2)}].
\ee
If and only (\ref{2int9})  holds, the system has a 2Dl  space of solutions $V$.
Since $V=\ {\rm constant}$ is always a solution,  (\ref{2int9}) is necessary and sufficient for the existence
of a nonzero 1-parameter potential system. In this case we can prescribe the values $V$, $V_2$ at any regular point ${\bf x}_0$; there will exist a
unique $V({\bf x})$ taking  these values.

\subsection{Free triplets}
A {\it 2nd
  order classical free triplet} is a  2D system without potential,
${\cal H}_0=\frac{p_1^2+p_2^2}{\lambda(x,y)}$
and with a basis of
 3  functionally independent second-order
constants of the motion 
$ {\cal L}_{(s)}=\sum_{i,j=1}^2a^{ij}_{(s)} p_ip_j,\  a^{ij}_{(s)}=a^{ji}_{(s)}$, $s=1,2,3$,
${\cal L}_{(3)}={\cal H}_0$. Since the duals of these constants of the motion are 2nd order Killing tensors, the spaces associated with
free triplets can be characterized as 2D manifolds that admit 3 functionally independent 2nd order Killing tensors. All such manifolds were classified by Koenigs
\cite{Koenigs,KKMW} who showed that the
possibilities were constant curvature spaces [each admitting 3 linearly independent 1st order Killing vectors], 4 Darboux spaces,
 [each admitting a single Killing vector] and 11 Koenigs spaces [each admitting no Killing vectors].   Since the vectors $\{ {\bf h_{(s)}}\}$,
${\bf h_{(s)}}^{\rm tr}(x,y,z)=\left( a^{11}_{(s)}, a^{12}_{(s)}, a^{22}_{(s)}\right)$
form a linearly independent set, there exist unique $3\times 3$ matrices ${\cal C}^{(j)}$ such that
$ \partial_{x_j}{\bf h}_{(s)}={\cal C}^{(j )}{\bf h}_{(s)}$,  $j,s=1,2$.  By linearity,  any element
 ${\cal L}=\sum_{i,j=1}^2a^{ij} p_ip_j$ of the space of 2nd order symmetries spanned by the basis triplet is characterized by matrix equations
\be\label{2int6}
\partial_{x_j}{\bf h}={\cal C}^{(j )}{\bf h}\qquad  j=1,2,\quad {\bf h}^{\rm tr}(x,y,z)=\left( a^{11},a^{12},a^{22}\right).
\ee
In particular, at any regular point ${\bf x}_0$ we can arbitrarily choose the value of the 3-vector ${\bf h}_0$ and solve (\ref{2int6}) to
 find the unique symmetry $\cal L$ of
 ${\cal H}_0$  such that  ${\bf h}({\bf x}_0)={\bf h}_0$. A normalization condition for the ${\cal C}^{(j)}$:  (\ref{2int6}) is valid for
$a^{11}=a^{22}={1}/{\lambda}, a^{12}=0$, i.e., for ${\cal H}_0$. Note that since the $\cal L$ are Killing tensors, equations (\ref{2int5}) must be compatible with the Killing equations
(\ref{Killingeqns}).  Also,   integrability conditions hold:
\be\label{2int7} {\cal C}^{(2)}_1-{\cal C}^{(1)}_2={\cal C}^{(1)}{\cal C}^{(2)}-{\cal C}^{(2)}{\cal C}^{(1)}\equiv [{\cal C}^{(1)},{\cal C}^{(2)}].
\ee

It is clear from equations (\ref{2Dsym}) that the restriction of a 2D 2nd order nondegenerate superintegrable system with all parameters  equal to 0
is a free triplet. However  the converse doesn't hold. We determine  necessary and sufficient conditions that a free system extends to a system with
nondegenerate potential.

A first step is a more detailed characterization of the matrices ${\cal C}^{(i)}$ for a free system.
 From the Killing equations (\ref{Killingeqns}) we obtain the conditions
\[{\cal C}_{11}^{(1)}=-G_1,\ {\cal C}_{12}^{(1)}=-G_2,\ {\cal C}_{13}^{(1)}=0,\
{\cal C}_{31}^{(2)}=0,\ {\cal C}_{32}^{(2)}=-G_1,\ {\cal C}_{33}^{(2)}=-G_2,\]
\[ 2{\cal C}_{21}^{(1)}+{\cal C}_{11}^{(2)}=0,\ 2{\cal C}_{22}^{(1)}+{\cal C}_{12}^{(2)}=-G_1,\ 2{\cal C}_{23}^{(1)}+{\cal C}_{13}^{(2)}=-G_2,\]
\[ 2{\cal C}_{21}^{(2)}+{\cal C}_{31}^{(1)}=-G_1,\ 2{\cal C}_{22}^{(2)}+{\cal C}_{32}^{(1)}=-G_2,\ 2{\cal C}_{23}^{(2)}+{\cal C}_{33}^{(1)}=-G_1.\]
From the requirement that ${\cal H}_0$ satisfies (\ref{2int6}) we obtain the conditions
\[ {\cal C}_{11}^{(1)}+{\cal C}_{13}^{(1)}=-G_1,\ {\cal C}_{21}^{(1)}+{\cal C}_{23}^{(1)}=0,\ {\cal C}_{31}^{(1)}+{\cal C}_{33}^{(1)}=-G_1,\]
\[ {\cal C}_{11}^{(2)}+{\cal C}_{13}^{(2)}=-G_2,\ {\cal C}_{21}^{(2)}+{\cal C}_{23}^{(2)}=0,\ {\cal C}_{31}^{(2)}+{\cal C}_{33}^{(2)}=-G_2.\]
Solving these equations we find
\[ {\cal C}^{(1)}=\left(\ba{ccc}-G_1,&-G_2,&0\\
 -\frac12{\cal C}^{(2)}_{11},&-\frac12G_1-\frac12{\cal C}^{(2)}_{12},&\frac12{\cal C}^{(2)}_{11}\\
-G_1-2{\cal C}^{(2)}_{21},& -G_2-2{\cal C}^{(2)}_{22},&2{\cal C}^{(2)}_{21}\ea\right),
\
 {\cal C}^{(2)}=\left(\ba{ccc}{\cal C}^{(2)}_{11},&{\cal C}^{(2)}_{12},&-G_2-{\cal C}^{(2)}_{11}\\
 {\cal C}^{(2)}_{21},&{\cal C}^{(2)}_{22},&-{\cal C}^{(2)}_{21}\\
0,& -G_1,&-G_2\ea\right),
\]
with the 4 functions $ {\cal C}^{(2)}_{11},\ {\cal C}^{(2)}_{12},\ {\cal C}^{(2)}_{21},\ {\cal C}^{(2)}_{22}$ free.
If we define the functions $A^{12},\ B^{12},\ A^{22},\ B^{22}$ by the requirement
\[ {\cal C}^{(2)}_{11}=-\frac23 G_2-\frac23 A^{12},\ {\cal C}^{(2)}_{12}=\frac13 G_1-\frac23 A^{22},\
 {\cal C}^{(2)}_{21}=-\frac13 G_1-\frac13B^{12},\ {\cal C}^{(2)}_{22}=-\frac23G_2-\frac13 B^{22},\]
then equations  (\ref{2int6}) agree with (\ref{2Dsym}). Thus, for a free system there always exist unique functions $A^{ij},B^{ij}$ such that equations (\ref{2Dsym}) hold.
Then necessary and sufficient conditions for extension to a  system with nondegenerate potential $V$ satisfying  equations   (\ref{canonicalequations}) are that 
conditions (\ref{2int3}) hold identically.

This analysis also extends, via restriction,  to superintegrable systems with degenerate potential. A free triplet  that corresponds to a degenerate superintegrable system is one
  that corresponds
 to a
nondegenerate system but  such  that one of the free generators can be chosen as a perfect square.
For these systems  conditions (\ref{2int9}) for the potential are satisfied identically.

Similarly, we define a {\it 2nd
  order quantum free triplet} as a  2D quantum system without potential,
$H_0=\frac{1}{\lambda({\bf x})}(\partial_{11}+\partial_{22})$,
and with a basis  of
 3  algebraically independent second-order
symmetry operators
\[L_k=\frac{1}{\lambda}\sum_{i,j=1}^2\partial_{i}(\lambda a^{ij}_{(k)}\partial_j)
({\bf x}),\ k=1,2,3,\  a^{ij}_{(k)}=a^{ji}_{(k)},\  L_3=H_0\]
There is a 1-1 relationship between classical and quantum free triplets. 

\section{Superintegrable systems and enveloping algebras} \label{section4.1}
Every 2D nondegenerate or degenerate superintegrable system is  St\"ackel equivalent to a superintegrable system on a constant curvature space \cite{KKM20042}. Thus we study  free triplets on  flat space and the  complex sphere,  taking advantage of the fact that the symmetries can be identified with 2nd order elements  in
the enveloping algebras of $e(2,\C)$ or $so(3,\C)$. Then,  conditions (\ref{2int7}) are satisfied.

 If we have a degenerate superintegrable system  and turn off the
potential then we  have a free degenerate superintegrable system in the sense that the Poisson brackets of the free generators determine a
degenerate quadratic algebra (without parameters). We will show, conversely, that every free triplet that forms
degenerate quadratic algebra is the restriction of a superintegrable system with degenerate potential.
We classify free triplet  systems that are 2nd order in  the enveloping algebras of $e(2,\C)$ and $o(3,\C)$ and which determine a degenerate quadratic algebra.
In the classification we identify systems that are equivalent under the adjoint
action of the corresponding  Lie group. We will also identify each system
 with the  superintegrable system with potential whose potential-free  terms agree with it. For this we use the classification of constant
curvature  systems in \cite{KKMP} with $E_{3'}$ added in \cite{Kress2007}. 
We start with flat space and consider free triplets in the  $e(2,\C)$ enveloping algebra.

\subsection{Degenerate superintegrable systems from  $e(2,\C)$  (8 systems)}\label{free2c}
We use the classical realization for $e(2,\C)$ with basis $p_1,p_2, {\cal J}=xp_2-yp_1$, 
and Hamiltonian ${\cal H}=p_1^2+p_2^2$. We classify all possible free degenerate superintegrable systems in
the enveloping algebra of $e(2,\C)$, up to conjugacy, modulo $\cal H$.
It turns out that each such system is the restriction of a degenerate flat space superintegrable system with potential;  the relationship is 1-1. We write $\tilde E_n$
as the free system that is the restriction of superintegrable system $E_n$.   Up to conjugacy under the action of
$e(2,\C)$, the possible choices for the 1st order generator $\cal X$ are: ${\cal X}={\cal J},\  p_1,\ p_1+ip_2$. We give some details
for the first case and then just list the results.

\medskip\noindent
 We first  choose ${\cal X}={\cal J}$. We need to find 2nd order elements ${\cal L}_1,\ {\cal L}_2$ of the enveloping algebra such that
$\{ {\cal X}^2,{\cal L}_1,{\cal L}_2,{\cal H}\}$ is linearly independent and such that
$\{ {\cal X},{\cal L}_1,{\cal L}_2,{\cal H}\}$ define a degenerate quadratic algebra. The most general choice for ${\cal L}_1$ is
$ {\cal L}_1=a_1Jp_1+a_2Jp_2+a_3 p_1^2+a_4p_1p_2$.
 Case 1: suppose $a_1\ne 0$ so we can take $a_1=1$. By a rotation, leaving $\cal J$ fixed, we can assume that ether $a_2=0$ or $a_2=i$.
We first consider: $a_2=0$. We can translate in $x$ to achieve $a_4=0$ and in $y$ to achieve $a_3=0$. Then for ${\cal L}_2$ we can
 take
$ {\cal L}_2=b_1Jp_2+b_2p_1^2+b_3p_1p_2$.
In order for these choices to generate a superintegrable system we must have
\be\label{XL1}\{{\cal X},{\cal L}_1\}=C_1{\cal L}_1+C_2{\cal L}_2+C_3{\cal H}+C_4{\cal X}^2+C_5,\ee
\be\label{XL2} \{{\cal X},{\cal L}_2\}=D_1{\cal L}_1+D_2{\cal L}_2+D_3{\cal H}+D_4{\cal X}^2+D_5,\ee
\be\label{L1L2} \{{\cal L}_1,{\cal L}_2\}=E_1{\cal L}_1X+E_2{\cal L}_2{\cal X}+E_3{\cal H}{\cal X}+E_4{\cal X}^3+E_5{\cal X}\ee 
\[G=c_1 {\cal L}_1^2+c_2{\cal L}_2^2+c_3{\cal H}^2+c_4{\cal L}_1{\cal  L}_2+c_5{\cal H}{\cal L}_1+c_6{\cal H}{\cal L}_2+c_7{\cal X}^4
+c_8{\cal X}^2{\cal L}_1+c_9{\cal X}^2 {\cal L}_2\]
\be\label{Geqn}+c_{10}{\cal H}{\cal X}^2+c_{11}{\cal L}_1+c_{12}{\cal L}_2+c_{13}{\cal H}+c_{14}{\cal X}^2+c_{15}\equiv 0,\ee
for some constants $A_j,C_j,E_j,c_j$ where the $c_j$ are not all 0. In ${\cal L}_2$ we assume first that $b_1\ne 0$ and normalize to $B_1=1$.
Then substituting into  equation (\ref{XL1}) and equating coefficients  of powers of $p_j$, $x$ and $y$ on both sides of the identity. We get easily that
$ C_1=C_3=C_4=C_5=0,\   C_2=-1,\  b_2=b_3=0$,
so there is no solution unless ${\cal L}_1={\cal X}p_1,L_2={\cal X}p_2$.  All remaining conditions are satisfied.
Now consider the case $b_1=0$ and assume $b_2=1$. This time equation (\ref{XL1}) cannot be solved, so this case is impossible. Next we assume $b_1=b_2=0,\ B_3=1$.
Again,  equation (\ref{XL1}) cannot be solved, so this case is also impossible. 
Now we consider the possibility $a_1=1, a_2=i$.  By translating in $y$
we can achieve $a_3=0$. Going step-by-step, we take $b_1=1$. Then we can satisfy (\ref{XL1})
only if $a_4=0$, in which case we have $\{{\cal X},{\cal L}_1\}={\cal L}_1$. Going further we now substitute this result  into equation
(\ref{XL2}) and equate coefficients.
We find a solution only if $b_2=b_3=0$, but now the space spanned by ${\cal L}_1,{\cal L}_2$ is the same as that spanned by ${\cal J}p_1,{\cal J}p_2$, already listed. 
This finishes Case 1. For Case 2 we can take $a_1=0$, $a_2=1$, and find no solutions.This finishes Case 2. 
For case 3 we take $a_1=a_2=0, a_3=1$. Here there is a solution. Having demonstrated the step-by-step approach,  we now merely list the results.
\begin{enumerate} 
\item $\tilde E_{18}$:\
 $ {\cal H}=p_1^2+p_2^2,\ {\cal X}= {\cal J},\ {\cal L}_1={\cal J}p_1,\ {\cal L}_2={\cal J}p_2,$ \hfill\break
$ {\rm Casimir:}\ -\frac12( {\cal L}_1^2+{\cal L}_2^2-{\cal H}{\cal X}^2)=0,\ {\rm potential}:\  V=\frac{\alpha}{\sqrt{x^2+y^2}}$,

\item $\tilde E_3$: $ {\cal H}=p_1^2+p_2^2,\ {\cal X}= {\cal J},\ {\cal L}_1=p_1^2,\ {\cal L}_2=p_1p_2$, \hfill\break
${\rm Casimir:}\ -{\cal L}_2^2-{\cal L}_1({\cal L}_1-{\cal H})=0, \ {\rm potential}:\  V=\alpha(x^2+y^2)$.

 \item $\tilde E_6$:\ ${\cal H}=p_1^2+p_2^2,\ {\cal X}= p_1,\ {\cal L}_1={\cal J}^2,\ {\cal L}_2={\cal J}p_2$,\hfill\break
$ {\rm Casimir:}\ {\cal L}_1{\cal X}^2+{\cal L}_2^2-{\cal H}{\cal L}_1=0, \ {\rm potential}:\  V=\frac{\alpha}{x^2}$,

 \item $\tilde E_5$:\ $ {\cal H}=p_1^2+p_2^2,\ {\cal X}= p_1,\ {\cal L}_1={\cal J}{p_1},\ {\cal L}_2=p_2{p_1}$,\hfill\break
$ {\rm Casimir:}\ \frac12({\cal L}_2^2+{\cal X}^4-{\cal H}{\cal X}^2)=0, \ {\rm potential}:\  V=\alpha x$.

  \item $\tilde E_{12}$:\ ${\cal H}=p_1^2+p_2^2,\ {\cal X}= p_1+ip_2,\ {\cal L}_1={\cal J}^2+(p_1-ip_2)^2,\ {\cal L}_2={\cal J}(p_1+ip_2)$,\hfill\break
${\rm Casimir:}\ i({\cal L}_1{\cal X}^2-{\cal L}_2^2-{\cal H}^2)=0,  \ {\rm potential}:\  V=\frac{\alpha (x+iy)}{\sqrt{(x+iy)^2+c^2}}$,

\item $\tilde E_{14}$:\ $ {\cal H}=p_1^2+p_2^2,\ {\cal X}= p_1+ip_2,\ {\cal L}_1={\cal J}^2,\ {\cal L}_2={\cal J}(p_1+ip_2)$,\hfill\break
$ {\rm Casimir:}\ i({\cal L}_1{\cal X}^2-{\cal L}_2^2)=0,  \ {\rm potential}:\  V=\frac{\alpha}{(x+iy)^2}$,

 \item $\tilde E_4$:\ $ {\cal H}=p_1^2+p_2^2,\ {\cal X}= p_1+ip_2,\ {\cal L}_1={\cal J}(p_1+ip_2),\ {\cal L}_2=(p_1-ip_2)^2$,\hfill\break
$ {\rm Casimir:}\ -i({\cal L}_2{\cal X}^2-{\cal H}^2)=0,  \ {\rm potential}:\  V=\alpha(x+iy)$,

\item  $\tilde E_{13}$:\ $ {\cal H}=p_1^2+p_2^2,\ {\cal X}= p_1+ip_2,\ {\cal L}_1={\cal J}(p_1+ip_2),\ {\cal L}_2=(p_1-ip_2){\cal J}$,\hfill\break
$ {\rm Casimir:}\ i({\cal L}_1{\cal H}-{\cal L}_2{\cal X}^2)=0,  \ {\rm potential}:\  V=\frac{\alpha}{\sqrt{ x+iy}}$.
\end{enumerate}

\subsection{Degenerate quadratic algebras from  $o(3,\C)$ (3 systems)}\label{o3Cdeg}
We use the classical realization for $o(3,\C)$ with basis ${\cal J}_1=yp_3-zp_2,\ {\cal J}_2=zp_1-xp_3,\ {\cal J}_3=xp_2-yp_1$, 
and Hamiltonian ${\cal H}={\cal J}_1^2+{\cal J}_2^2+{\cal J}_3^2$.  We classify  the possible systems up to conjugacy with respect to $O(3,\C)$ group
actions and modulo $\cal H$
using the same step-by-step procedure as in Section \ref{free2c}, and merely list the results.
Up to conjugacy, the  choices for $\cal X$ are ${\cal J}_3$,  ${\cal J}_1+i{\cal J}_2$.
\begin{enumerate}
\item  $\tilde S_6$:\ $ {\cal H}={\cal J}_1^2+{\cal J}_2^2+{\cal J}_3^2,\ {\cal X}= {\cal J}_3,\ {\cal L}_1={\cal J}_3{\cal J}_1,\ {\cal L}_2={\cal J}_3{\cal J}_2$,\hfill\break
$ {\rm Casimir:}\ -\frac12({\cal L}_1^2+{\cal L}_2^2+{\cal X}^2({\cal X}^2-{\cal H}))=0,  \ {\rm potential}:\  V=\frac{\alpha z}{\sqrt{x^2+y^2}}$,

\item $\tilde S_3$:\ $ {\cal H}={\cal J}_1^2+{\cal J}_2^2+{\cal J}_3^2,\ {\cal X}= {\cal J}_3,\ {\cal L}_1=({\cal J}_1+i{\cal J}_2)^2,\
 {\cal L}_2=({\cal J}_1-i{\cal J}_2)^2$,\hfill\break
${\rm Casimir:}\ -2i( ({\cal H}-{\cal X}^2)^2-{\cal L}_1{\cal L}_2)=0,   \ {\rm potential}:\  V=\frac{\alpha }{z^2}$,
\item $\tilde S_5$:\ ${\cal H}={\cal J}_1^2+{\cal J}_2^2+{\cal J}_3^2,\  {\cal X}= {\cal J}_1+i{\cal J}_2,\ {\cal L}_1={\cal J}_3^2,\ 
{\cal L}_2=({\cal J}_1+i{\cal J}_2){\cal J}_3$,\hfill\break
$ {\rm Casimir:}\ -i({\cal L}_2^2-{\cal X}^2{\cal L}_1)=0,   \ {\rm potential}:\  V=\frac{\alpha }{(x+iy)^2}$.
\end{enumerate}
\subsection{Nondegenerate quadratic algebras from  $e(2,\C)$ (12  plus 1)} \label{nondegE}
We  use the realization for $e(2,\C)$ with basis listed in Section \ref{free2c}.
An alternate basis is ${\cal J}, p_1+ip_2, p_1-ip_2$.
We classify systems, mod $\cal H$, up to conjugacy with respect to the group $E(2,\C)$, including inversions and reflections.
There are 8 conjugacy classes of 2nd order elements in the enveloping algebra, mod $\cal H$, with representatives
\be\label{e2conj} {\cal J}^2,\quad  p_1^2,\quad (p_1+ip_2)^2,\quad p_2{\cal J},\quad (p_1+ip_2){\cal J},\quad {\cal J}^2+ap_1^2,\ a\ne0,\ee
\[ {\cal J}^2+(p_1+ip_2)^2,\quad 2(p_1+ip_2){\cal J}+(p_1-ip_2)^2.\]
A general 2nd order element in the enveloping algebra, mod $\cal H$, can be written as
$a_1{\cal J}^2+a_2p_1{\cal J}+a_3p_2{\cal J}+a_4p_1^2+a_5p_1p_2$.

\medskip\noindent
{\bf 1st case:} We choose ${\cal L}_1={\cal J}^2$ and try to determine the possibilities for ${\cal L}_2$, up to conjugacy under $E(2,\C)$, such that ${\cal L}_1,{\cal L}_2,{\cal H}$ generate a quadratic algebra.
(As we go through the cases step-by-step, we ignore systems that  have already been exhibited in earlier steps.)
In general
$ {\cal L}_2=a_2p_1{\cal J}+a_3p_2{\cal J}+a_4p_1^2+a_5p_1p_2$
and $a_2,a_3,a_4,a_5$  are to be determined. Here
${\cal R}=\{{\cal L}_1,{\cal L}_2\}=-2a_2p_2{\cal J}+2a_3p_1{\cal J}-4a_4p_1p_2{\cal J}+2a_5(2p_1^2-{\cal H}){\cal J}$.
We must require that
$ {\cal R}^2= b_1 {\cal L}_1^3 + b_2 {\cal L}_2^3 + b_3 {\cal H}^3 + b_4 {\cal L}_1^2{\cal L}_2 + b_5 {\cal L}_1{\cal L}_2^2
+ b_6 {\cal H}{\cal L}_1{\cal L}_2 +b_7{\cal  H}{\cal  L}_1^2 + b_8 {\cal H}{\cal  L}_2^2 + b_9 {\cal H}^2{\cal L}_1 + b_{10}{\cal  H}^2{\cal  L}_2$,
for some constants $b_1,\cdots, b_{10}$. We substitute our expressions for ${\cal L}_1$ and ${\cal L}_2$ into ${\cal R}^2$ and equate coefficients of powers of $p_1,p_2,x,y$
on both sides of the resulting equation. These yields a system of equations for the parameters $a_j,b_k$, polynomial in the $a_j$ and linear in the $b_k$.
The step-by-step procedure to solve for the parameters is similar to that demonstrated
earlier for degenerate systems. Once a solution is obtained we check that it extends to a superintegrable system with potential by using the generators to compute the functions $A^{ij},B^{ij}$ and then verifying directly that these functions satisfy the integrability conditions
(\ref{2int3}). Then we identify the associated nondegenerate superintegrable system from the classification in \cite{KKMP}. 
 We list the results, eliminating duplicates and exhibiting the 3-parameter potentials of the associated nonfree superintegrable systems.
\begin{enumerate}
 \item $\tilde E_{16}$:\quad ${\cal L}_1={\cal J}^2$, ${\cal L}_2=p_1{\cal J}$, ${\cal R}^2=4{\cal L}_1({\cal L}_1{\cal H}-{\cal L}_2^2)$. \hfill\break
$V=\frac{1}{\sqrt{x^2+y^2}}(\alpha+\frac{\beta}{y+\sqrt{x^2+y^2}}+\frac{\gamma}{y-\sqrt{x^2+y^2}})$,

\item $\tilde E_{17}$:\quad ${\cal L}_1={\cal J}^2$, ${\cal L}_2=(p_1+ip_2){\cal J}$, ${\cal R}^2=-4{\cal L}_1{\cal L}_2^2$, \hfill\break
$V=\frac{\alpha}{\sqrt{x^2+y^2}}+\frac{\beta}{(x+iy)^2}+\frac{\gamma}{(x+iy)\sqrt{x^2+y^2}}$,

\item $\tilde E_1$:\quad ${\cal L}_1={\cal J}^2$, $ {\cal L}_2=p_1^2$, ${\cal R}^2=16{\cal L}_1{\cal L}_2({\cal H}-{\cal L}_2)$, \hfill\break
$V=\alpha(x^2+y^2)+\frac{\beta}{x^2}+\frac{\gamma}{y^2}$,

\item $\tilde E_8$\quad ${\cal L}_1={\cal J}^2$, ${\cal L}_2=(p_1+ip_2)^2$, ${\cal R}^2=-16{\cal L}_1{\cal L}_2^2$, \hfill\break
$V=\frac{\alpha (x-iy) }{(x+iy)^3}+\frac{\beta}{(x+iy)^2}+\gamma(x^2+y^2)$,

\item $\tilde E_{3'}$:\quad ${\cal L}_1=p_1^2$, ${\cal L}_2=p_1p_2$, ${\cal R}^2=0$, \hfill\break
$V=\alpha(x^2+y^2)+\beta x+\gamma y$,

\item $\tilde E_2$:\quad ${\cal L}_1=p_2^2$, ${\cal L}_2=p_2{\cal J}$, ${\cal R}^2=4{\cal L}_1^2({\cal H}-{\cal L}_1)$.\hfill\break
$V=\alpha(4x^2+y^2)+\beta x+\frac{\gamma}{ y^2}$,

 \item $\tilde E_7$:\quad ${\cal L}_1=(p_1+ip_2)^2$, ${\cal L}_2={\cal J}^2+\frac{b}{2}(p_1-ip_2)^2,\ b\ne 0$, 
${\cal R}^2=-16{\cal L}_1^2{\cal L}_2+16a{\cal L}_1{\cal H}^2$, \hfill\break
$V=\frac{\alpha(x+iy)}{\sqrt{(x+iy)^2-b}}+\frac{\beta (x-iy)}{\sqrt{(x+iy)^2-b}\ \left(x+iy+\sqrt{(x+iy)^2-b}\ \right)^2}+\gamma (x^2+y^2)$,

\item $\tilde E_9$:\quad ${\cal L}_1=(p_1+ip_2)^2$, ${\cal L}_2=p_1{\cal J}$, ${\cal R}^2=-2{\cal L}_1({2\cal L}_1+{\cal H})^2$, \hfill\break
$V=\frac{\alpha}{\sqrt{x+iy}}+\beta y+\frac{\gamma (x+2iy)}{\sqrt{x+iy}}$,

\item $\tilde E_{11}$:\quad ${\cal L}_1=(p_1+ip_2)^2$, ${\cal L}_2=(p_1-ip_2){\cal J}$, ${\cal R}^2=-4{\cal L}_1{\cal H}^2$, \hfill\break
$V=\alpha(x-iy)+\frac{\beta (x-iy)}{\sqrt{x+iy}}+\frac{\gamma }{\sqrt{x+iy}}$,

\item $\tilde E_{10}$:\quad ${\cal L}_1=(p_1-ip_2)^2$, ${\cal L}_2=4i(p_1-ip_2){\cal J}+(p_1+ip_2)^2$, ${\cal R}^2=64{\cal L}_1^3$, \hfill\break
$V=\alpha(x-iy)+\beta (x+iy-\frac32(x-iy)^2)+\gamma(x^2+y^2-\frac12(x-iy)^3)$,

\item ${\tilde E_{15}}$:\quad ${\cal L}_1=(p_1-ip_2)^2$, ${\cal L}_2=i(p_1-ip_2){\cal J}$, ${\cal R}^2=4{\cal L}_1^3$,
\hfill\break
$V=f(x-iy)$, where $f$ is arbitrary. The exceptional case, characterized by  the fact that generators 
${\cal L}_1,{\cal L}_2,{\cal H}$ are functionally linearly dependent, \cite{KKMP,KKM20041}. 
This  quadratic algebra is isomorphic, but not conjugate, to $\tilde E_{10}$ and doesn't correspond to a 
nondegenerate  superintegrable system. 

 \item $\tilde E_{20}$:\quad ${\cal L}_1=p_2{\cal J}$, ${\cal L}_2=p_1{\cal J}$, ${\cal R}^2={\cal H}({\cal L}_1^2+{\cal L}_2^2)$,
 \hfill\break
$V=\frac{1}{\sqrt{x^2+y^2}}\left(\alpha+\beta \sqrt{x+\sqrt{x^2+y^2}}+\gamma \sqrt{x-\sqrt{x^2+y^2}}\right)$,

 \item $\tilde E_{19}$:\quad ${\cal L}_1= (p_1+ip_2){\cal J}$, ${\cal L}_2={\cal J}^2+(p_1-ip_2)^2$, 
 ${\cal R}^2=-4{\cal L}_2({\cal L}_1^2+{\cal H}^2)$,  \hfill\break
$V=\frac{\alpha(x+iy)}{\sqrt{(x+iy)^2-4}}+\frac{\beta}{\sqrt{(x-iy)(x+iy+2)}}+\frac{\gamma}{\sqrt{(x-iy)(x+iy-2)}}$.

\end{enumerate}

\subsection{Nondegenerate quadratic algebras from $o(3,\C)$ enveloping algebra (6 systems)} \label{nondegS}
We make use of  the classical realization for $o(3,\C)$ given in Section \ref{o3Cdeg}.  
We classify the possible systems up to conjugacy with respect to $O(3,\C)$ group actions and modulo $\cal H$.
There are 5 conjugacy classes of 2nd order elements in the enveloping algebra, mod $\cal H$, with representatives
\be\label{e2conj1} {\cal J}_3^2,\quad  {\cal J}_1^2+a{\cal J}_2^2,\ (a\ne 0,\pm 1,\ |a|\le 1),\quad ({\cal J}_1+i{\cal J}_2)^2,\ee
\[ ({\cal J}_1+i{\cal J}_2)^2+{\cal  J}_3^2,\quad {\cal J}_3({\cal J}_1+i{\cal J}_2).\]
A general 2nd order element in the enveloping algebra , mod $\cal H$ can be written as
$a_1{\cal J}_1^2+a_2{\cal J}_2^2+a_3{\cal J}_1{\cal J}_2+a_4{\cal J}_1{\cal J}_3+a_5{\cal J}_2J_3$.
An alternate expression is
$ A_1({\cal J}_1+i{\cal J}_2)^2+A_2({\cal J}_1-i{\cal J}_2)^2+A_3 {\cal J}_3^2+A_4({\cal J}_1+i{\cal J}_2){\cal J}_3+A_5({\cal J}_1-i{\cal J}_2){\cal J}_3$.

 We list the results, eliminating duplicates and exhibiting the 3-parameter 
potentials of the associated nonfree superintegrable systems.
\begin{enumerate}
\item $\tilde S_9$:\quad  ${\cal L}_1={\cal J}_3^2$,
 ${\cal L}_2={\cal J}_1^2$, ${\cal R}^2=-16{\cal L}_1^2{\cal L}_2-16{\cal L}_1{\cal L}_2^2+16{\cal L}_1{\cal L}_2{\cal H}$,  \hfill\break
$V=\frac{\alpha}{x^2}+\frac{\beta}{y^2}+\frac{\gamma}{z^2}$,

\item $\tilde S_4$:\quad  ${\cal L}_1={\cal J}_3^2$, ${\cal L}_2=({\cal J}_1+i{\cal J}_2){\cal J}_3$, 
${\cal R}^2=-4{\cal L}_1{\cal L}_2^2$,  \hfill\break
$V=\frac{\alpha}{(x+iy)^2}+\frac{\beta z}{\sqrt{x^2+y^2}}+\frac{\gamma}{(x+iy)\sqrt{x^2+y^2}}$,

\item $\tilde S_7$: ${\cal L}_1={\cal J}_3^2$, ${\cal L}_2={\cal J}_1{\cal J}_3$, 
 ${\cal R}^2=-4{\cal L}_1^3-4{\cal L}_2^2{\cal L}_1+4{\cal L}_1^2{\cal H}$,  \hfill\break
$V=\frac{\alpha z}{\sqrt{x^2+y^2}}+\frac{\beta x}{y^2\sqrt{x^2+y^2}}+\frac{\gamma}{y^2}$,

\item $\tilde S_8$: ${\cal L}_1={\cal J}_2({\cal J}_2+i{\cal J}_1)$, ${\cal L}_2={\cal J}_2{\cal J}_3$, 
 ${\cal R}^2=-2{\cal L}_1^3+2{\cal L}_1{\cal L}_2^2+{\cal L}_1^2{\cal H}-{\cal L}_2^2{\cal H}$,  \hfill\break
$V=\frac{\alpha y}{\sqrt{x^2+z^2}}+\frac{\beta (y+ix+z)}{\sqrt{(y+ix)(z+ix)}}+\frac{\gamma(y+ix-z)}{\sqrt{(y+ix)(z-ix)}}$,

\item  $\tilde S_2$:\quad ${\cal L}_1= ({\cal J}_1+i{\cal J}_2)^2$, ${\cal L}_2={\cal J}_3^2$,
 ${\cal R}^2=-16{\cal L}_1^2{\cal L}_2$,  \hfill\break
$V=\frac{\alpha }{z^2}+\frac{\beta }{(x+iy)^2}+\frac{\gamma(x-iy)}{(x+iy)^3}$,

\item $\tilde S_1$:\  ${\cal L}_1= ({\cal J}_1+i{\cal J}_2){\cal J}_3$, ${\cal L}_2=({\cal J}_1+i{\cal J}_2)^2$, 
${\cal R}^2=-4{\cal L}_2^3$,  \hfill\break
$V=\frac{\alpha }{(x+iy)^2}+\frac{\beta z}{(x+iy)^2}+\frac{\gamma(1-4z^2)}{(x+iy)^4}$,
\end{enumerate}

\subsection{The closure  theorems}
There are, up to conjugacy, 8 degenerate and 13 nondegenerate quadratic algebras in the enveloping algebra of $e(2,\C)$, and these match 1-1 with the restrictions of the 8 degenerate, 12 nondegenerate and 1 exceptional superintegrable systems on complex flat space, also classified up to conjugacy. There are, up to conjugacy,   3 degenerate and 6 nondegenerate quadratic algebras in the enveloping algebra of $o(3,\C)$, and these match 1-1 with the restrictions of the 3 degenerate and 6 nondegenerate superintegrable systems on the complex 2-sphere.
  Thus:
\begin{theorem} A classical free triplet on a constant curvature space extends to a superintegrable system if and only if it forms a
free quadratic algebra, degenerate or nondegenerate.
\end{theorem}
The main message that follows from this result is that we have found purely algebraic conditions on constant curvature spaces that replace the complicated
analytic integrability conditions (\ref{2int3}) or (\ref{2int7})  for
extension to a superintegrable system. 

There is an analogous result for quantum free systems and quantum superintegrable systems. Indeed, If we have a  nondegenerate quantum
superintegrable system  and turn off the
potential then we will have a free nondegenerate superintegrable system in the sense that the commutators  of the free generators will determine a
nondegenerate quadratic algebra. Conversely, every quantum free triplet system for which the algebra formed from the
generators closes to a
nondegenerate  quadratic algebra is the restriction of a superintegrable system with nondegenerate potential (or the exceptional case $E_{15}$). Indeed, since the highest order
derivative terms in the  commutator agree
with the highest order polynomial terms in the Poisson bracket, every free quantum nondegenerate quadratic algebra uniquely determines a
free classical nondegenerate
quadratic algebra. The classical quadratic algebras correspond 1-1 with classical superintegrable systems and these in turn correspond 1-1
with quantum superintegrable systems. There is a similar correspondence for degenerate quadratic algebras. Thus we have
\begin{theorem} A quantum  free triplet on a constant curvature space extends to a superintegrable system if and only if it forms a
free quantum quadratic algebra.
\end{theorem}
 In a forthcoming paper we will show that these theorems extend to all 2D superintegrable systems, including those on Darboux and Koenig spaces.

\subsection{Construction of superintegrable systems from free triplets}
Suppose we have a classical free triplet with basis
\[ {\cal L}_{(s)}=\sum_{i,j=1}^2a^{ij}_{(s)} p_ip_j\quad a^{ij}_{(s)}=a^{ji}_{(s)},\ s=1,2,3,\ {\cal L}_{(3)}={\cal H}_0=\frac{p_1^2+p_2^2}{\lambda(x,y)},\]
not ${\tilde E15}$, 
 that determines a free nondegenerate quadratic algebra, hence  a free nondegenerate superintegrable system.
Then the functions $A^{ij},B^{ij}$, (\ref{canoneqns}) expressed in terms of the Cartesian-like coordinates $(x,y)$, 
satisfy the integrability conditions  (\ref{2int3}) for the potential equations (\ref{canonicalequations})
and we are guaranteed  a 4-dimensional vector space of solutions $V$. Further, these 
equations  guarantee that the Bertrand-Darboux integrability conditions for equations
 $W^{(s)}_i=\lambda\sum_{j=1}^2 a^{ij}_{(s)}V_j$
are satisfied and we can compute the solutions $ W^{(s)}$, $W^{(3)}=V$, unique up to  additive constants, such that the constants of the motion
${\cal L}_{(s)} =\sum
a^{ij}_{(s)}p_ip_j+W^{(s)}$ define a nondegenerate superintegrable system. This system is guaranteed to satisfy a nondegenerate quadratic algebra with potential whose highest
order (potential-free) terms agree with the free quadratic algebra. Note that the functions $A^{ij},B^{ij}$ are defined independent of the basis chosen
for the free triplet, although, of course,
they do depend upon the particular coordinates chosen.
Similarly, there is an associated  quantum free triplet
\[  L_s=\frac{1}{\lambda}\sum_{i,j=1}^2\partial_{i}(\lambda a^{ij}_{(s)}\partial_j)
,\ s=1,2,3,\  L_3=H_0=\frac{1}{\lambda({\bf x})}(\partial_{11}+\partial_{22}),\]
that defines  a free nondegenerate quantum quadratic algebra with potential. The functions $W^{(s)}$ are the same as before.

There is an analogous construction of degenerate superintegrable systems with potential from  free triplets that generate a free  quadratic algebras,
but are such that one generator say, ${\cal L}_1={\cal X}^2$ is a perfect square. Then the system with its generator added determines a free degenerate quadratic algebra.
The functions $A^{ij},B^{ij}$ are defined from the free triplet
and ${\cal X}=\xi_1p_1+\xi_2p_2$. The equations for the potential are
\be \label{degpot1} V_2 = C^2V_1,\  V_{22}=V_{11} + C^{22}V_1,\  V_{12} = C^{12}V_1,\ee
where
 $ C^2(x_1,x_2)=-\frac{\xi_1}{\xi_2},\ C^{22}(x_1,x_2)=A^{22} -\frac{\xi_1}{\xi_2}B^{22},\ C^{12}(x_1,x_2)= A^{12}-\frac{\xi_1}{\xi_2}B^{12}$.
Since the system determines a quadratic algebra, the integrability conditions for the potential equations (\ref{degpot1}) are satisfied identically and the solution
space is 2-dimensional. The general solution takes the form $ V= a_1V^{(0)}+a_2$ where $a_1,a_2$ are constant coefficients. This defines the degenerate
superintegrable system. The extension to the quantum case is obvious.

\begin{example} $E_1$: From \S \ref{nondegE} we have the classical free system
${\cal L}_1={\cal J}^2$, $ {\cal L}_2=p_1^2$, ${\cal R}^2=16{\cal L}_1{\cal L}_2({\cal H}-{\cal L}_2)$,
Using Cartesian coordinates $x_1=x,x_2=y$ we find
$A^{12}=0,\quad A^{22}=\frac{3}{x},\quad B^{12}=0,\quad B^{22}=-\frac{3}{y}$.
The general solution of the potential equations is
$V=a_1(x^2+y^2)+\frac{a_2}{x^2}+\frac{a_3}{y^2}+a_4$.
Setting $a_4=0$ we find that the  induced classical system is
\be {\cal H}=p_1^2+p_2^2+a_1(x^2+y^2)+\frac{a_2}{x^2}+\frac{a_3}{y^2},\ee
\[ {\cal  L}_1= (xp_2-yp_1)^2 +{a_2y^2}/{x^2}+{a_3x^2}/{y^2},\
         {\cal L}_2=p_1^2+a_1x^2+{a_2}/{x^2}.\]            
 The induced Casimir is
$ {\cal R}^2=16\left({\cal L}_1{\cal L}_2{\cal H}-{\cal L}_1{\cal L}_2^2-(a_2+a_3){\cal L}_2^2-a_2{\cal H}^2\right.$\hfill\break 
$\left. +2a_2{\cal L}_2{\cal H}-a_1{\cal L}_1^2+4a_1a_2a_3\right)$.
The quantum system is defined by
\bea \ba{rl} H=&\partial_x^2+\partial_y^2+a_1(x^2+y^2)+\frac{a_2}{x^2}+{a_3}/{y^2},\
  L_2=\partial_x^2+a_1x^2+{a_2}/{x^2},
\\
L_1=& (x\partial_y-y\partial_x)^2 +{a_2y^2}/{x^2}+{a_3x^2}/{y^2},
                \ea
         \label{E1Operators}\eea
 The induced Casimir is
$
 R^2=\frac{8}{3}\left(\{L_1,L_2, H\}-\{L_2, L_2,L_1\}\right)-(16a_2+12)H^2$
 $-\left(\frac{176}{3}+16a_2+16a_3\right)L_2^2 -16a_1L_1^2 +\left(\frac{176}3+32a_2\right)L_2H+\frac{176a_1}{3}L_1$\hfill\break
$ -\frac{16a_1}3\left(12a_2a_3+9a_2+9a_3+2\right)$.
\end{example}
\begin{example} $S_9$: From \S \ref{nondegS} we have the classical free systems
 ${\cal L}_1={\cal J}_3^2$, ${\cal L}_2={\cal J}_1^2$, ${\cal R}^2=-16{\cal L}_1^2{\cal L}_2-16{\cal L}_1{\cal L}_2^2+16{\cal L}_1{\cal L}_2{\cal H}$.
The structure equations are more symmetrical if we choose a new basis symmetry ${\cal L}_3={\cal J}_2^2$ in place of $\cal H$.
Using  coordinates  $x_1=\psi,x_2=\phi$ where $s_1=\frac{\cos\phi}{\cosh\psi},\ s_2=\frac{\sin\phi}{\cosh\psi},\ s_3=\tanh\psi$,
 $s_1^2+s_2^2+s_3^2=1$,
the Hamiltonian is $H=\cosh^2\psi(p_\psi^2+p_\phi^2),\quad \lambda=\frac{1}{\cosh^2\psi}$,
\[A^{12}=0,\quad A^{22}=\frac{3\cosh^2\psi-\sinh^2\psi}{\sinh\psi\cosh\psi},\quad B^{12}=2\frac{\sinh\psi}{\cosh\psi},\
 B^{22}=-3\frac{(\cos^2\phi-\sin^2\phi)}{\sin\phi\cos\phi}.\]
The general potential is
$V=\frac{a_1}{s_1^2}+\frac{a_2}{s_2^2}+\frac{a_3}{s_3^2}+a_4$.
Setting $a_4=0$ we find the classical symmetries.
The induced classical $S9$ system has a basis of symmetries
\be\label{comS9} {\cal L}_2={\cal
  J}_1^2+a_2\frac{s_3^2}{s_2^2}+a_3\frac{s_2^2}{s_3^2},\ {\cal L}_3={\cal
  J}_2^2+a_3\frac{s_1^2}{s_3^2}+a_1\frac{s_3^2}{s_1^2},\
{\cal L}_1={\cal
  J}_3^2+a_1\frac{s_2^2}{s_1^2}+a_2\frac{s_1^2}{s_2^2},
\ee
where ${\cal H}={\cal L}_1+{\cal L}_2+{\cal L}_3+a_1+a_2+a_3$. The classical Casimir is
\[ {\cal R}^2=16{\cal L}_1{\cal L}_2{\cal L}_3-16a_2{\cal L}_3^2-16a_3{\cal L}_1^2-16a_1{\cal L}_2^2+64a_1a_2a_3.\]
The quantum superintegrable system is defined as
\bea \ba{rl} H=&J_1^2+J_2^2+J_3^2+\frac{a_1}{s_1^2}+\frac{a_2}{s_2^2}+\frac{a_3}{s_3^2},\
 L_1= J_3^2 +\frac{a_2 s_1^2}{s_2^2}+\frac{a_1^2s_2^2}{s_1^2},\\
        L_2=& J_1^2+\frac{a_3 s_2^2}{s_3^2}+\frac{a_2 s_3^2}{s_2^2},\
        L_3= J_2^2+\frac{a_1 s_3^2}{s_1^2}+\frac{a_3s_1^2}{s_3^2},\ea \label{S9Operators}\eea
$H=L_1+L_2+L_3+a_1+a_2+a_3$.
The quantum Casimir is
\[  R^2=\frac83\{L_1,L_2,L_3\} -(16a_3+12)L_1^2 -(16a_1+12)L_2^2  -(16a_2+12)L_3^2\]
\[+\frac{52}{3}(\{L_1,L_2\}+\{L_2,L_3\}+\{L_3,L_1\})+ \frac13(16+176a_3)L_1\]
\[+\frac13(16+176a_1)L_2 + \frac13(16+176a_2)L_3 +\frac{32}{3}(a_1+a_2+a_3)\]
\[+48(a_1a_2+a_2a_3+a_3a_1)+64a_1a_2a_3.\]
\end{example}
\begin{example} $S_3$: This is a restriction of system $\tilde S_9$ in the preceding example and we use the same notation. We set ${\cal L}_1={\cal X}^2$, ${\cal X}=p_\phi$,
 $x_1=\psi,x_2=\phi$. We have
$C^{2}=0$, $ C^{22}=\frac{3\cosh^2\psi-\sinh^2\psi}{\sinh\psi\cosh\psi}$, $C^{12}=0$,
so
$ V_2=0,\ V_{11}+\frac{3\cosh^2\psi-\sinh^2\psi}{\sinh\psi\cosh\psi}V_1=0$.
The general potential is
$V=\frac{a_3}{s_3^2}+a_4$. 
The induced classical $S3$ system has a basis of symmetries and Casimir relation
\[ {\cal H}'={\cal J}_1^2+{\cal J}_2^2+{\cal J}_3^2+\frac{a_3}{s_3^2},\
 {\cal L}'_1={\cal J}_1^2+a_3\frac{s_2^2}{s_3^2},\  {\cal L}_2'={\cal J}_1{\cal J}_2-a_3\frac{s_1s_2}{s_3^2},\
{\cal X}={\cal J}_3,\]
\be\label{classcasimirS3}{{\cal L}_1'}^2 + {{\cal L}_2'}^2- {\cal L}_1' {\cal H}'+ {\cal L}_1' {\cal X}^2+ a_3{\cal X}^2+ a_3{\cal L}_1=0.
\ee
The quantum superintegrable system is defined as
\[ H=J_1^2+J_2^2+J_3^2+\frac{a_3}{s_3^2},\
 X=J_3, \ L_1=J_1^2 + \frac{a_3 s_2^2}{s_3^2}, \
L_2= \frac12(J_1J_2+J_2J_1)-\frac{a_3 s_1s_2}{s_3^2}.\]
The Casimir is
$
\{L_1, X^2\}+2L_1^2+2L_2^2-2L_1H+\frac{5+4a_3}{2}X^2-2aL_1-a_3=0$.
\end{example}

\section{Contractions of superintegrable systems}\label{contractions} Suppose we have a nondegenerate quantum superintegrable system with
generators $H,L_1,L_2$ and structure
 equations (\ref{Casimir1}), defining a quadratic algebra $Q$.
If we make a change of basis to new generators ${\tilde H},{\tilde L_1}, {\tilde L_2}$ and parameters ${\tilde a_1},{\tilde a_2},
 {\tilde a_3}$ such that
\be\label{nondegcontraction}
\left(\begin{array}{c}
{\tilde L_1}\\
{\tilde L_2} \\
{\tilde H}
\end{array}\right)
=\left(\begin{array}{ccc}
A_{1,1} & A_{1,2}&A_{1,3} \\
A_{2,1}&A_{2,2} &A_{2,3}  \\
0 &0 &A_{3,3}
\end{array}\right)
\left(\begin{array}{c}
L_1\\
L_2 \\
H
\end{array}\right)+
\left(\begin{array}{ccc}
B_{1,1} & B_{1,2}&B_{1,3} \\
B_{2,1}&B_{2,2} &B_{2,3}  \\
B_{3,1} &B_{3,2} &B_{3,3}
\end{array}\right)
\left(\begin{array}{c}
a_1\\
a_2 \\
a_3
\end{array}\right),\ee
\[ \left(\begin{array}{c}
{\tilde a_1}\\
{\tilde a_2} \\
{\tilde a_3}
\end{array}\right)
=\left(\begin{array}{ccc}
C_{1,1} & C_{1,2}&C_{1,3} \\
C_{2,1}&C_{2,2} &C_{2,3}  \\
C_{3,1} &C_{3,2} &C_{3,3}
\end{array}\right)
\left(\begin{array}{c}
a_1\\
a_2 \\
a_3
\end{array}\right)
\]
for some $3\times 3$ constant matrices $A=(A_{i,j}),B,C$ such that $\det A \cdot \det C\ne 0$,  we will have the same
system with new structure equations  of the form (\ref{Casimir1}) for ${\tilde R}=[{\tilde L_1},{\tilde L_2}]$, $[{\tilde L_j},
{\tilde R}]$,  ${\tilde R}^2$,  but with transformed structure constants. (Strictly speaking, since the space of potentials is 4-dimensional,
we should have a term $a_4$ in the above expressions. However, normally, this term can be absorbed into $H$. Also, we could add constant terms to each of the symmetries
${\tilde H}, {\tilde L_j}$ but we shall restrict ourselves to this class of basis changes here.)
 We choose a continuous 1-parameter family of basis transformation matrices $A(\epsilon),B(\epsilon),C(\epsilon)$,
$0<\epsilon\le 1$ such
that $A(1)=C(1)$ is the identity matrix, $B(1)=0$ and  $\det A(\epsilon)\ne 0$, $\det C(\epsilon)\ne 0$. Now
suppose as $\epsilon\to 0$ the basis change becomes singular, (i.e., the limits of $A,B,C$ either do not exist or,
if they exist
 do not satisfy $\det A(0)\det C(0)\ne 0$) but the structure equations involving $A(\epsilon),B(\epsilon),C(\epsilon)$, go to a limit,
  defining a new quadratic algebra $Q'$. We call $Q'$ a {\it contraction} of $Q$ in analogy with Lie algebra contractions \cite{Wigner}. 
We can also define contractions of free superintegrable systems in an obvious manner from (\ref{nondegcontraction}): Just set $a_1=a_2=a_3=0$ and  $B=C=0$.

For a degenerate superintegrable system with generators $H,X, L_1,L_2$ and structure equations (\ref{structure2}),(\ref{Casimir2}),
defining a quadratic algebra $Q$,
 a change of basis to new generators ${\tilde H},{\tilde X},{\tilde L_1}, {\tilde L_2}$ and
parameter ${\tilde a}$ such that ${\tilde a} = Ca$, and
\[
\left(\begin{array}{c}
{\tilde L_1}\\
{\tilde L_2} \\
{\tilde H}\\
{\tilde X}\\
{\tilde X}^2
\end{array}\right)
=\left(\begin{array}{ccccc}
A_{1,1} & A_{1,2}&A_{1,3}&0&A_1 \\
A_{2,1}&A_{2,2} &A_{2,3}&0 &A_2 \\
0 &0 &A_{3,3}&0&0\\
0&0&0&A_{4,4}&0\\
0&0&0&0&A_{4,4}^2
\end{array}\right)
\left(\begin{array}{c}
L_1\\
L_2 \\
H\\
X\\
X^2
\end{array}\right)+
\left(\begin{array}{c}
B_1 \\
B_2  \\
B_3\\
0\\
0
\end{array}\right)
a
\]
for some $4\times 4$   matrix $A=(A_{i,j})$, in the upper left-hand corner, with $\det A\ne 0$, complex 4-vectors $A'=(A_i)$, $B$ and constant $C\ne 0$ yields the
same superintegrable
system with new structure equations  of the form (\ref{structure2}),(\ref{Casimir2}) for $[{\tilde X},{\tilde L_j}]$,
$[{\tilde L_1},{\tilde L_2}]$, and $\tilde G=0$,  but with transformed structure constants.  
 Suppose we choose a continuous 1-parameter family of basis transformation
matrices $(A(\epsilon),A'(\epsilon)),B(\epsilon),C(\epsilon)$, $0<\epsilon\le 1$ such that $A(1)$ is the identity matrix, $A'(1)=B(1)=0$, $C(1)=1$, and
$\det A(\epsilon)\ne 0$, $ C(\epsilon)\ne 0$. Now suppose as $\epsilon\to 0$ the basis change becomes singular but that the structure equations
involving $A(\epsilon),A'(\epsilon),B(\epsilon),C(\epsilon)$ go to a finite limit, thus defining a new quadratic algebra $Q'$.
We call $Q'$ a {\it contraction} of $Q$.
 Contractions of  free degenerate superintegrable systems
are defined in an analogous manner: Set $a=0$, $B=0$. 
There are analogous definitions of contractions for classical systems.

\subsection{Lie algebra contractions of $o(3,\C)$ and $e(2,\C)$}
In general, the classification of possible contractions of quadratic algebras is very complex, but for quadratic algebras associated with
systems on constant curvature spaces,  there is a class of contractions with important 
physical/geometrical significance
that can easily be classified: contractions induced from Lie algebra contractions. In \cite{Wigner} In\"on\"u and Wigner defined a family of contractions of Lie algebras,
with special emphasis on the symmetry algebras of constant curvature spaces: the Wigner-In\"on\"u contractions. Later a larger class of contractions was studied,
 so-called natural contractions, \cite{Saletan}. We recall the definition of natural (quantum) contraction.
Let $(A; [ ; ]_A)$, $(B; [ ; ]_B)$ be two complex Lie algebras. We say
$B$ is a {\it contraction} of $A$ if for every $\epsilon\in (0; 1]$ there exists a linear invertible
map $t_\epsilon : B\to A$ such that for every $X, Y\in B$, $\lim_{\epsilon\to 0}t_\epsilon^{-1}[t_\epsilon X,t_\epsilon Y]_A
= [X, Y ]_B$.
Thus,  as $\epsilon\to 0$ the 1-parameter family of basis transformations can become nonsingular but the structure constants go to a finite limit.
There is an analogous definition for classical contractions.
For  Lie algebras $e(2,\C)$ and $so(3,\C)$ the contractions  have all been classified up to conjugacy,
\cite{Conaster,NP, WW}. We first list these contractions and their physical implementations, then show that they induce contractions of  free
nondegenerate and degenerate classical quadratic algebras
associated with constant curvature spaces  and, ultimately, contractions of the nondegenerate and degenerate (classical and quantum)
superintegrable systems with potential. We omit contractions to the abelian algebra and the identity contractions,  irrelevant for our purposes.

 We start with $e(2,\C)$ and  use the classical realization  with basis $p_1,p_2,{\cal J}=xp_2-yp_1$ and Hamiltonian ${\cal H}=p_1^2+p_2^2$.

\be\label{e2contractions}{\rm \bf Contractions\ of}\  e(2,\C):\ee
\begin{enumerate}
\item $ \{{\cal J}',p_1',p_2'\}=\{ {\cal J},\ \epsilon p_1,\ \epsilon p_2\}: \  e(2,\C),$\hfill  \break
$ {\rm coordinate\ implementation}\ x'=\frac{x}{\epsilon},y'=\frac{y}{\epsilon},$
\item $\{  {\cal J}',\  p_1'+ip_2',\ p_1'-ip_2'\}=\{{\cal J},\ \epsilon(p_1+ip_2),\ p_1-ip_2\}\ :\ e(2,\C),$\hfill\break
${\rm coordinate\ implementation}\ x'+iy'=x+iy,x'-iy'=\frac{x-iy}{\epsilon},$
\item
 $ \{{\cal J}',p_1',p_2'\}=\{ {\cal J}+\frac{p_1}{\epsilon},\ p_1,\ p_2\}:\   e(2,\C),$\hfill\break
${\rm coordinate\ implementation}\ x'=x,y'=y-\frac{1}{\epsilon},$
 \item $\{{\cal J}',p_1',p_2'\}=\{ {\cal J}+\frac{p_1+ip_2}{\epsilon},\ p_1,\  p_2\}:\  e(2,\C),$\hfill\break
${\rm coordinate\ implementation}\ x'=x+\frac{i}{\epsilon},y'=y-\frac{1}{\epsilon}.$

These last two contraction types  can be combined, even including  different powers of $\epsilon$. A relevant example is 
\be\label{combinedcontraction}  \{{\cal J}',p_1',p_2'\}=\{ {\cal J}+\frac{p_1+ip_2}{\epsilon}+\frac{p_1-ip_2}{\sqrt{\epsilon}},\ p_1,\  p_2\}:\  e(2,\C),\ee
$\qquad\qquad  {\rm coordinate\ implementation}\ x'=x+\frac{i}{\epsilon}-\frac{i}{\sqrt{\epsilon}},y'=y-\frac{1}{\epsilon}-\frac{1}{\sqrt{\epsilon}}.$
\item $ \{{\cal J}',p_1',p_2'\}=\{\epsilon {\cal J},\ p_1,\ \epsilon p_2\}\ : \ {\rm Heisenberg\ algebra},$\hfill\break
${\rm coordinate\ implementation}\ x'=x,y'=\frac{y}{\epsilon}, {\cal J}'=x'p_2'$,
\end{enumerate}
\bigskip

We use the classical realization for $o(3,\C)$ acting on the 2-sphere, with basis ${\cal J}_1=s_2p_3-s_3p_2,\ {\cal J}_2=s_3p_1-s_1p_3,\ {\cal J}_3=s_1p_2-s_2p_1$, 
Hamiltonian ${\cal H}={\cal J}_1^2+{\cal J}_2^2+{\cal J}_3^2$. Here $\sum_{j=1}^3s_j^2=1$ and restriction to the sphere gives
$s_1p_1+s_2p_2+s_3p_3=0$.
\be\label{o3contractions}{\rm \bf Contractions\ of} \ o(3,\C):\ee
\begin{enumerate}
\item $\{{\cal J}_1',{\cal J}_2',{\cal J}_3'\}=\{ \epsilon {\cal J}_1,\ \epsilon {\cal J}_2,\  {\cal J}_3\}:\  e(2,\C),$\hfill\break
${\rm coordinate\ implementation}\ x={s_1}/{\epsilon},y={s_2}/{\epsilon}, s_3\approx 1, {\cal J}={\cal J}_3,$
\item $ \{ {\cal J}_1'+i{\cal J}_2',\  {\cal J}_1'-i{\cal J}_2',\  {\cal J}_3'\}=\{ {\cal J}_1+i{\cal J}_2,\  \epsilon({\cal J}_1-i{\cal J}_2),\  {\cal J}_3\}:\  e(2,\C),$\hfill\break
${\rm coordinate\ implementation}\ s_1+is_2=\epsilon z,\ s_1-is_2= {\bar z},\ s_3\approx 1,$\hfill\break
$  {\cal J}_3'=i(zp_z-{\bar z}p_{\bar z}),$
${\cal J}_1'+i{\cal J}_2' =2ip_{\bar z},\ {\cal J}_1'-i{\cal J}_2'=-2ip_z,$
 \item $  \{{\cal J}'_1+i{\cal J}_2',{\cal J}_1'-i{\cal J}_2',{\cal J}_3'\}=\{ \epsilon({\cal J}_1+i{\cal J}_2),\frac{{\cal J}_1-i{\cal J}_2}{\epsilon},{\cal J}_3\}:\   o(3,\C),$\hfill\break
${\rm coordinate\ implementation}\
 s_1'=\frac{\epsilon+\epsilon^{-1}}{2}s_1+i\frac{\epsilon-\epsilon^{-1}}{2}s_2,$\hfill\break
$ s_2'=-i\frac{\epsilon-\epsilon^{-1}}{2}s_1+\frac{\epsilon+\epsilon^{-1}}{2}s_2,\  s_3'=s_3,$
\item $\{ {\cal J}_1'+i{\cal J}_2',\  {\cal J}_1'-i{\cal J}_2',\  {\cal J}_3'\}=\{\epsilon( {\cal J}_1+i{\cal J}_2),\  {\cal J}_1-i{\cal J}_2,\ \epsilon {\cal J}_3\}:\ {\rm Heisenberg\ algebra},$ \hfill\break
${\rm coordinate\ implementation}\ s_1=\frac{\cos\phi}{\cosh\psi},\  s_2=\frac{\sin\phi}{\cosh\psi},\  s_3=\frac{\sinh\psi}{\cosh\psi},$\hfill\break
${\rm we\ set}\ \phi=\epsilon\theta-i\ln \sqrt{\epsilon},\ \psi=\xi\sqrt{\epsilon},\ {\rm to\ get\ }$\hfill\break
$  {\cal J}_3'=p_\theta,\ {\cal J}_1'+i{\cal J}_2'=\xi p_\theta-ip_\xi,\ {\cal J}_1'-i{\cal J}_2'=\xi p_\theta+ip_\xi.$
\end{enumerate}

\subsection{Quadratic enveloping algebra contractions from Lie algebra contractions}
Note that once we choose a basis for a Lie algebra $A$,  its enveloping algebra is uniquely determined by the structure constants.
Structure relations in the enveloping
algebra are continuous functions of the structure constants. Thus a contraction of one Lie algebra $A$ to another, $B$  induces a  contraction of the
corresponding  enveloping algebras of $A$ and $B$. In the case of $e(2,\C)$, $o(3,\C)$, free quadratic algebras constructed in the enveloping algebras will contract
to free quadratic algebras generated by the target Lie algebras.

Consider only  4 contractions of $e(2,\C)$ to itself and 1 to the Heisenberg algebra. Each of the first 4  when applied to a free nondegenerate or
degenerate quadratic
algebra $\tilde E_j$
will contract to a a quadratic algebra $\tilde E_k$ where $k$ may be distinct from $j$. The last contraction will also lead to a quadratic algebra which we call {\it singular}
because the new Hamiltonian will be degenerate. We do not classify these singular systems but they are of physical and mathematical interest. Of the 4 nontrivial contractions of $o(3,\C)$, 1 takes $o(3,\C)$ to itself (so $\tilde S_j$ to $\tilde S_k$),
2 take it to $e(2,\C)$ (so $\tilde S_j$ to $\tilde E_k$   and 1 to the Heisenberg algebra (so $\tilde S_j$ to a singular system).
\begin{example}
{\bf  ${\tilde E_1}\to {\tilde E_8}$:} Use $\{  {\cal J}',\  p_1'+ip_2',\ p_1'-ip_2'\}=\{  {\cal J},\  \epsilon(p_1+ip_2),\ p_1-ip_2\}$. $ {\cal H}'=\epsilon {\cal H}=(p_1'+ip'_2)(p_1'-ip_2')$, ${\cal L}'_1=({\cal J}')^2={\cal L}_1$,
 ${\cal L}'_2=4\epsilon^2{\cal L}_2=(p_1'+ip_2')^2$.
\end{example}

\begin{example}$\tilde E_1\to \tilde  E_1$.  Use $\{{\cal J}',p_1',p_2'\}=\{{\cal J},\epsilon p_1,\epsilon p_2\}$. $ {\cal L}_1'={\cal J}^2={\cal L}_1$, $
 {\cal L}_2'=\epsilon^2 p_1^2=\epsilon^2 {\cal L}_2$,
${\cal H}'=\epsilon^2(p_1^2+p_2^2)= \epsilon^ 2 {\cal H}$.
\end{example}
\begin{example} $\tilde E_1\to \ {\rm Heisenberg}$. Use $\{{\cal J}',p_1',p_2'\}=\{\epsilon {\cal J},p_1,\epsilon p_2\}$.
$ {\cal L}_1'=\epsilon^2 {\cal L}_1={{\cal J}'}^2={x'}^2{p_2'}^2$,
 ${\cal L}_2'= {\cal L}_2={p_1'}^2$,
${\cal H}'=\epsilon^2({\cal H}-{\cal L}_2)={p_2'}^2$.
Structure relations:
$ {\cal R}=\{{\cal L}_1',{\cal L}_2'\},\  {\cal R}^2=4{p_1'}^2{p_2'}^4=4{\cal L}_1'{{\cal H}'}^2$.
\end{example}

\begin{example}
$\tilde S_3\to \tilde S_3$. $\{{\cal J}_1'+i{\cal J}_2',{\cal J}_1'-i{\cal J}_2',{\cal J}_3'\}=\{ \epsilon({\cal J}_1+i{\cal J}_2),({\cal J}_1-i {\cal J}_2)/\epsilon, {\cal J}_3\}$.
$
{\cal X}'= {\cal X}={\cal J}_3,\ {\cal H}'={{\cal J}_1'}^2+{{\cal J}_2'}^2+{{\cal J}_3'}^2=  {\cal H}$,
 ${\cal L}_1'= ({\cal J}_1'+i{\cal J}_2')^2=4i\epsilon^2{\cal L}_2$,
${\cal L}_2'= ({\cal J}_1'-i{\cal J}_2')^2=\frac{2}{\epsilon^2}({\cal L}_1-i{\cal L}_2-\frac12 {\cal H}+\frac12 {\cal X}^2)$.
\end{example}
We list the  contractions in  tables. For $e(2,\C)$ the relevant contractions are (\ref{e2contractions}): 
\be\label{Table2} \ba{lllllll}&&& e(2,\C)& {\rm Contractions}&(\rm degenerate)&\\
\hline\\
&&e(2)\to e(2),{\bf 1}&e(2)\to e(2),{\bf 2}&e(2)\to e(2),{\bf 3}&e(2)\to e(2),{\bf 4}&e(2)\to\ {\rm Heisenberg}\\
\hline\\
{\tilde E}_3&:&{\tilde E}_3&{\tilde E}_3&{\tilde E}_5&{\tilde E}_{4}&{\cal L}_1'{{\cal H}'}={{\cal L}_2'}^2\\
{\tilde E}_4 &:&{\tilde E}_4 &{\tilde E}_4 &{\tilde E}_4 &{\tilde E}_{4}&{{\cal H}'}={{\cal X}'}^2\\
{\tilde E}_{5}&:&{\tilde E}_{5}&{\tilde E}_{4}&{\tilde E}_{5}&{\tilde E}_{5}&{{\cal L}'}^2={{\cal X}'}^2{\cal H}'\\
{\tilde E}_6&:&{\tilde E}_6&{\tilde E}_{14}&{\tilde E}_5&{\tilde E}_5&{{\cal L}_2'}^2={\cal L}_1'{\cal H}'\\
{\tilde E}_{12}&:&{\tilde E}_{4}&{\tilde E}_{12}&{\tilde E}_{4}&{\tilde E}_{12}&{{\cal L}_2'}^2={\cal L}_1'{\cal H}'+{{\cal H}'}^2\\
{\tilde E}_{13}&:&{\tilde E}_{13}& {\tilde E}_{13}&{\tilde E}_4&{\tilde E}_{13}&{\cal L}_1'{\cal L}_3'={\cal H}{\cal L}_2' \\
{\tilde E}_{14}&:&{\tilde E}_{14}&{\tilde E}_{14}&{\tilde E}_4&{\tilde E}_{14}&{{\cal L}_2'}^2={{\cal L}_1'}{\cal H}'\\
{\tilde E}_{18}&:&{\tilde E}_{18}&{\tilde E}_{18}&{\tilde E}_{5}&{\tilde E}_{13}&{{\cal L}_2'}^2={{\cal X}'}^2{\cal H}'
\ea
\ee

\be\label{Table1} \ba{lllllll}&&& e(2,\C)& {\rm Contractions}&(\rm nondegenerate)&\\
\hline\\
&&e(2)\to e(2),{\bf 1}&e(2)\to e(2),{\bf 2}&e(2)\to e(2),{\bf 3}&e(2)\to e(2),{\bf 4}&e(2)\to\ {\rm Heisenberg}\\
\hline\\
{\tilde E}_1&:&{\tilde E}_1&{\tilde E}_8&{\tilde E}_2&{\tilde E}_{3'}&{{\cal R}'}^2={\cal L}_1'{{\cal H}'}^2\\
{\tilde E}_8 &:&{\tilde E}_8 &{\tilde E}_8 &{\tilde E}_3' &{\tilde E}_{15}, {\tilde E}_{10}&{{\cal R}'}^2={\cal L}_1'{{\cal H}'}^2\\
{\tilde E}_{10}&:&{\tilde E}_{3'}&{\tilde E}_{3'}&{\tilde E}_{3'}&{\tilde E}_{10}&{{\cal R}'}^2={{\cal H}'}^3\\
{\tilde E}_2&:&{\tilde E}_2&{\tilde E}_{15}&{\tilde E}_2&{\tilde E}_{3'}&{{\cal R}'}^2={\cal L}_1'{{\cal H}'}^2\\
{\tilde E}_{3'}&:&{\tilde E}_{3'}&{\tilde E}_{3'}&{\tilde E}_{3'}&{\tilde E}_{3'}&{{\cal R}'}^2=0\\
{\tilde E}_{16}&:&{\tilde E}_{16}&{\tilde E}_{17}&{\tilde E}_2&{\tilde E}_{15}, {\tilde E}_{10}&{{\cal R}'}^2=4{{\cal L}_1'}^2{\cal H}'\\
{\tilde E}_7&:&{\tilde E}_{3'}&{\tilde E}_7&{\tilde E}_{3'}&{\tilde E}_{15}, {\tilde E}_{10}&{{\cal R}'}^2=4({\cal L}_1'+a{\cal H}'){{\cal H}'}^2\\
{\tilde E}_{17}&:&{\tilde E}_{17}&{\tilde E}_{17}&{\tilde E}_{3'}&{\tilde E}_{15}, {\tilde E}_{10}&{{\cal R}'}^2=0\\
{\tilde E}_{19}&:&{\tilde E}_{11}&{\tilde E}_{19}&{\tilde E}_{3'}&{\tilde E}_{15}, {\tilde E}_{10}&{{\cal R}'}^2=0\\
{\tilde E}_{11}&:&{\tilde E}_{11}&{\tilde E}_{11}&{\tilde E}_{11}&{\tilde E}_{11}&{{\cal R}'}^2={{\cal H}'}^3\\
{\tilde E}_9&:&{\tilde E}_9&{\tilde E}_{15},{\tilde E}_{11}&{\tilde E}_{3'}&{\tilde E}_9&{{\cal R}'}^2={{\cal L}_1'}^2{\cal H}'\\
{\tilde E}_{20}&:&{\tilde E}_{20}&{\tilde E}_{20}&{\tilde E}_{3'}&{\tilde E}_{11}&{{\cal R}'}^2={{\cal L}_1'}^2{\cal H}'
\ea
\ee
Note: For the ${\tilde E_7}\to {\tilde E_{10}}$, ${\tilde E_8}\to {\tilde E_{10}}$,  ${\tilde E_{16}}\to {\tilde E_{10}}$, ${\tilde E_{17}}\to {\tilde E_{10}}$ and ${\tilde E_{19}}\to {\tilde E_{10}}$ contractions we use (\ref{combinedcontraction}). For ${\tilde E_{10}}\to {\tilde E_{3'}}$, case {\bf 3} we use the composite contraction $\{{\cal J}',p_1',p_2'\}=\{{\cal J}+\frac{p_1+p_2}{\epsilon},p_1,p_2\}$.

The relevant $o(3,\C)$ contractions are, (\ref{o3contractions}):

\be\label{Table4} \ba{llllll}&& o(3,\C)& {\rm Contractions}&(\rm degenerate)&\\
\hline\\
&&o(3)\to e(2),{\bf 1}&o(3)\to e(2),{\bf 2}&o(3)\to o(3),{\bf 3}&o(3)\to\ {\rm Heisenberg}\\
\hline\\
{\tilde S}_3&:&{\tilde E}_3&{\tilde E}_3&{\tilde S}_{3}&{{\cal X}'}^2{{\cal H}'}={{\cal L}_2'}^2\\
{\tilde S}_5 &:&{\tilde E}_{14}&{\tilde E}_{14}&{\tilde S}_5&{{\cal X}'}^2{{\cal H}'}={{\cal L}_2'}^2\\
{\tilde S}_{6}&:&{\tilde E}_{18}&{\tilde E}_{18}&{\tilde S}_{6}&{{\cal X}'}^2={{\cal H}'}\ea
\ee

\be\label{Table3} \ba{llllll}&& o(3,\C)& {\rm Contractions}&(\rm nondegenerate)&\\
\hline\\
&&o(3)\to e(2),{\bf 1}&o(3)\to e(2),{\bf 2}&o(3)\to o(3),{\bf 3}&o(3)\to\ {\rm Heisenberg}\\
\hline\
{\tilde S}_9&:&{\tilde E}_1&{\tilde E}_8&{\tilde S}_{2}&{{\cal R}'}^2=-{{\cal L}_2'}^2{\cal H}'\\
{\tilde S}_4 &:&{\tilde E}_{17}&{\tilde E}_{17}&{\tilde S}_4&{{\cal R}'}^2=-4{{\cal L}_2'}^2{\cal H}'\\
{\tilde S}_{7}&:&{\tilde E}_{16}&{\tilde E}_{17}&{\tilde S}_4&{{\cal R}'}^2=-4{{\cal L}_2'}^2{\cal H}'\\
{\tilde S}_8&:&{\tilde E}_9&{\tilde E}_{11}&{\tilde S}_4&{{\cal R}'}^2=-4{{\cal L}_2'}^2{\cal H}'\\
{\tilde S}_{2}&:&{\tilde E}_8&{\tilde E}_8&{\tilde S}_{2}&{{\cal R}'}^2=-16{{\cal L}_2'}^2{\cal H}'\\
{\tilde S}_{1}&:&{\tilde E}_{11}&{\tilde E}_{11}&{\tilde S}_{1}&{{\cal R}'}^2=0
\ea
\ee

\subsection{Contractions/restrictions of free nondegenerate  systems to free degenerate ones}
These are not contractions in the standard sense. As we have shown in Section \ref{degstructure}, they arise through  the following mechanism.
 Suppose we take a particular basis of 2nd order generators ${\cal H}, {\cal L}_1,{\cal L}_2$  for the classical nondegenerate free  system such that
  the symmetry ${\cal L}_1$ is a perfect square:
${\cal L}_1={\cal X}^2$.
Then ${\cal X}$ will be a 1st order symmetry for ${\cal H}$, i.e.,  a Killing vector.
From the relation
$ {\cal R}=\{{\cal L}_1,{\cal L}_2\}=2{\cal X}\{{\cal X},{\cal L}_2\}$,
we see that ${\cal L}_3=\{{\cal X},{\cal L}_2\}$ is a 2nd order symmetry for ${\cal H}$ (which in most case turns out to be linearly
independent of the symmetries ${\cal H},{\cal L}_1,{\cal L}_2$ we already know). Then  we can factor $4{\cal X}^2$ from each term of
the identity ${\cal R}^2-{\cal F}=0$ to obtain the Casimir ${\cal G}=0$ for the contracted system, where ${\cal G}={\cal L}_3^2+\cdots$.
 In any case, we are guaranteed by theory that a 2nd order symmetry ${\cal L}_3$ exists such that ${\cal X}, {\cal L}_1,{\cal L}_2,{\cal L}_3,{\cal H}$ define a
unique free degenerate quadratic algebra. We give some examples:

\begin{enumerate}
 \item $\tilde S_9\to \tilde S_3$:\ 
In system $\tilde S_9$ we have  ${\cal L}_1={\cal J}^2_3, {\cal L}_2={\cal J}^2_1$.
We note that  ${\cal X}={\cal J}_3$, a Killing vector.
The Casimir for the original system is
\be\label{classS9a}{\cal R}^2=-16{\cal L}_1^2{\cal L}_2-16{\cal L}_1{\cal L}_2^2+16{\cal L}_1{\cal L}_2{\cal H} \ee
where ${\cal R}=\{{\cal L}_1,{\cal L}_2\}$. In the contracted system
we take   ${\cal L}_3=\{{\cal X},{\cal L}_2\}$. Setting ${\cal L}_3=2{\cal L}_2'$, ${\cal L}_2={\cal L}_1'$ we see that
 (\ref{classS9a}) reduces to the Casimir
$({\cal L}_1')^2 + ({\cal L}_2')^2- {\cal L}_1' {\cal H}+ {\cal L}_1' {\cal X}^2=0,
$
which can be identified with $\tilde S_3$.

\item $\tilde E_1\to \tilde E_3$:\ 
In system $\tilde E_1$:
${\cal L}_1={\cal J}^2,\  {\cal L}_2=p_1^2,\ {\cal R}^2=16{\cal L}_1{\cal L}_2({\cal H}-{\cal L}_2)$,
let ${\cal X}={\cal J}$.  Setting $\{{\cal X},{\cal L}_2\}=2{\cal L}_2'$, ${\cal L}_2={\cal L}_1'$,
we see that the Casimir for $\tilde E_1$ reduces to
${{\cal L}_1'}^2+{{\cal L}_2'}^2-{\cal L}_1'{\cal H}'=0$,
the structure equation  for $\tilde E_3$.

\item $\tilde E_8\to \tilde E_{14}$:\ 
In system $\tilde E_8$: ${\cal L}_1={\cal J}^2,\ {\cal L}_2=(p_1+ip_2)^2,\ {\cal R}^2=-16{\cal L}_1{\cal L}_2^2$,
let ${\cal X}=p_1+ip_2$. Setting ${\cal L}_1={\cal L}_1'$, $\{{\cal X},{\cal L}_1\}=2i{\cal L}_2'$
we find the Casimir
$-{{\cal L}_2'}^2+{\cal X}^2{\cal L}_1'=0$ for $\tilde E_{14}$.

\item $\tilde E_{10}\to \tilde E_4$:\ 
This case is less obvious.. In system  $\tilde E_{10}$:
${\cal L}_1=(p_1-ip_2)^2$, ${\cal L}_2=4i(p_1-ip_2){\cal J}+(p_1+ip_2)^2$, $ {\cal R}^2=64{\cal L}_1^3$,
we set ${\cal X}=p_1-ip_2$, ${\cal L}_2={\cal L}_1'$. Now the  Casimir restricts to
$4{\cal X}^2\{{\cal X},{\cal L}_1'\}^2=64 {\cal X}^6$,
or $\{{\cal X},{\cal L}_1'\}=\pm 4{\cal X}^2$.
We take the plus sign to be definite. It appears that this system closes on itself and doesn't give us a 4th generator.
 However, it follows from the analysis in \S \ref{free2c}  that there is a {\it unique} free degenerate quadratic algebra $E_4$ containing
the algebra generated by ${\cal X},{\cal H},{\cal L}_1'$, namely the one generated by ${\cal X},{\cal H},{\cal L}_1',{\cal L}_2'$ where ${\cal L}_2'=(p_1+ip_2)^2$.
\end{enumerate}

\bigskip
Contractions of free nondegenerate $e(2,\C)$ systems to degenerate systems:
\be\label{Table5} \ba{lllll}& e(2,\C)& {\rm Contractions}:&{\rm nondegenerate\ }\to & {\rm degenerate}\\
\hline\\
&&{\cal J}&p_1&p_1+ip_2\\
\hline\\
{\tilde E}_1&:&{\tilde E}_3&{\tilde E}_6&-\\
{\tilde E}_8 &:&{\tilde E}_3 &- &{\tilde E}_{14}\\
{\tilde E}_{10}&:&-&-&{\tilde E}_{14}\\
{\tilde E}_2&:&-&{\tilde E}_{6}, {\tilde E}_{5}&-\\
{\tilde E}_{3'}&:&{\tilde E}_{3}&{\tilde E}_{5}&-\\
{\tilde E}_{16}&:&{\tilde E}_{18}&-&-\\
{\tilde E}_7&:&{\tilde E}_{3}&-&{\tilde E}_{12}\\
{\tilde E}_{17}&:&{\tilde E}_{18}&-&-\\
{\tilde E}_{19}&:&-&-&-\\
{\tilde E}_{11}&:&-&-&{\tilde E}_{4}\\
{\tilde E}_9&:&-&-&{\tilde E}_{13}\\
{\tilde E}_{20}&:&-&-&-
\ea
\ee

\bigskip
Contractions of free nondegenerate $o(3,\C)$ systems to degenerate systems:
\be\label{Table6} \ba{llll}& o(3,\C)& {\rm Contractions}:&{\rm nondegenerate}\ \to {\rm degenerate}\\
\hline\\
&&{\cal J}_3&{\cal J}_1+i{\cal J}_2\\
\hline
{\tilde S}_9&:&{\tilde S}_3&-\\
{\tilde S}_4 &:&{\tilde S}_{6}&-\\
{\tilde S}_{7}&:&{\tilde S}_{6}&-\\
{\tilde S}_8&:&-&-\\
{\tilde S}_{2}&:&{\tilde S}_3&{\tilde S}_5\\
{\tilde S}_{1}&:&-&{\tilde S}_{5}
\ea
\ee

\subsection{Contractions of superintegrable systems with potential induced by free quadratic algebra  contractions}
Suppose we have a classical free triplet ${\cal H}^{(0)}, {\cal L}_1^{(0)}, {\cal L}_2^{(0)}$ that determines a nondegenerate quadratic algebra $Q^{(0)} $
and structure functions $A^{ij}({\bf x}), B^{ij}({\bf x})$ in some set of Cartesian-like coordinates $(x_1,x_2)$. Further, suppose this system contracts to another nondegenerate system
${{\cal H}'}^{(0)}, {{\cal L}'}_1^{(0)}, { {\cal L}'}_2^{(0)}$ with quadratic algebra ${Q'}^{(0)}$ via the mechanism described in the preceding sections. We show here that this contraction induces a contraction of the associated nondegenerate superintegrable system
${\cal H}={\cal H}^{(0)}+V$, ${\cal L}_1={\cal L}_1^{(0)}+W^{(1)}$,
 ${\cal L}_2={\cal L}_2^{(0)}+W^{(2)}$, $Q$ to
${\cal H}'={{\cal H}'}^{(0)}+V'$, ${\cal L}'_1={{\cal L}'}_1^{(0)}+{W^{(1)}}'$,
 ${\cal L}'_2={{\cal L}'}_2^{(0)}+{W^{(2)}}'$, $Q'$.
The point is that in  the contraction process the symmetries ${{\cal H}'}^{(0)}(\epsilon)$,
${{\cal L}'}_1^{(0)}(\epsilon)$,
${ {\cal L}'}_2^{(0)}(\epsilon)$
remain continuous functions of $\epsilon$, linearly independent as quadratic forms, and
 $\lim_{\epsilon\to 0} {{\cal H}'}^{(0)}(\epsilon)={{\cal H}'}^{(0)}$,
$\lim_{\epsilon\to 0} {{\cal L}'}^{(0)}_j(\epsilon)={{\cal L}'}^{(0)}_j$.
Thus the associated functions $A^{ij}(\epsilon), B^{ij}(\epsilon)$ will also be continuous functions of $\epsilon$ and
$\lim_{\epsilon\to 0}A^{ij}(\epsilon)={A'}^{ij}$, $\lim_{\epsilon\to 0}B^{ij}(\epsilon)={B'}^{ij}$. Similarly, the integrability conditions for the potential equations
\be\label{nondegpot2} \ba{lllll}
 V^{(\epsilon)}_{22}&=& V^{(\epsilon)}_{11}&+&A^{22}(\epsilon) V^{(\epsilon)}_1+B^{22}(\epsilon) V^{(\epsilon)}_2,\\
 V^{(\epsilon)}_{12}&=& &&A^{12}(\epsilon) V^{(\epsilon)}_1+B^{12}(\epsilon) V^{(\epsilon)}_2,\ea
\ee
will hold for each $\epsilon$ and in the limit. This means that the 4-dimensional solution space for the potentials $V$ will deform continuously into the 4-dimensional solution space for the potentials $V'$. Thus the target space of solutions $V'$ is uniquely determined by the free quadratic algebra contraction.

A similar argument using the functions $C^2,C^{22},C^{12}$ where
\be \label{degpot2} V_2 = C^2V_1,\  V_{22}=V_{11} + C^{22}V_1,\  V_{12} = C^{12}V_1,\ee
applies to contractions of free degenerate quadratic algebras. Again the 2-dimensional space of source potentials deforms continuously to the target space.
\begin{theorem} A Lie algebra contraction of the free quadratic algebra of a free triplet system to another such system induces a unique contraction relating  the associated superintegrable systems with potential.
\end{theorem}

There is an apparent lack of uniqueness in this procedure, since for a nondegenerate superintegrable system one typically chooses a basis $V^{(j)},\ j=1,\cdots,4$ for the potential space and expresses a general potential as $V=\sum_{j=1}^4a_jV^{(j)}$. Of course the choice of basis for the source system is arbitrary, as is the choice for the target system. Thus the structure equations for the quadratic algebras and the dependence $a_j(\epsilon)$ of the contraction constants on $\epsilon$ will vary depending on these choices. However, all such possibilities are related by a basis change matrix.

\begin{example}
We describe how a Lie algebra contraction induces the contraction of
$E_1$ to $E_2$, including the potential terms. Recall for $E_1$
in Cartesian coordinates $x_1=x,x_2=y$ we have
$H=p_x^2+p_y^2+V$,
\be\label{ABeqns4} A^{12}=0,\quad A^{22}=\frac{3}{x},\quad B^{12}=0,\quad B^{22}=-\frac{3}{y}.\ee
The general potential is
$ V=a_1(x^2+y^2)+\frac{a_2}{x^2}+\frac{a_3}{y^2}+a_4$.
For $E_2$ and using Cartesian coordinates $x_1=x',x_2=y'$ we have
$A'^{12}=0,\quad A'^{22}=0,\quad B'^{12}=0, \quad B'^{22}=-\frac{3}{y'}$.
Thus, the general potential for $E_2$ is 
$ V'=b_1(4x'^2+y'^2)+b_2x'+\frac{b_3}{y'^2}+b_4$.
In terms of these coordinates the  contraction is defined by
$x=x'+\frac{1}{\epsilon},\ y=y'$.
Substituting these values in (\ref{ABeqns4}) and going to the limit as $\epsilon\to 0$
we get $A'^{12}= A'^{22}=B'^{12}=0,\ B'^{22}=-\frac{3}{y'}$, the canonical equations for $E_2$.
In the limit the 4 dimensional space of potentials for $E_1$ must go to the 4 dimensional vector space for $E_2$.
However the chosen basis functions for the $E_1$ potential, 
$ x^2+y^2,\  \frac{1}{x^2},\  \frac{1}{y^2},\ 1$
will not go to a new basis in the limit; 1 basis function blows up and 1 basis function goes to 0.   One of the simplest choices of basis that avoids this problem   is
$V^{(1)}(\epsilon)=x^2+y^2+\frac{1}{\epsilon^4x^2}-\frac{1}{\epsilon^2}\to 4x'^2+y'^2$,
$V^{(2)}(\epsilon)=\frac{-1}{2\epsilon}( \frac{1}{\epsilon^2 x^2}-1)\to x',\ V^{(3)}(\epsilon)= \frac{1}{ y^2}\to \frac{1}{{y'}^2},\   V^{(4)}(\epsilon)=1\to 1$.
Thus if we set
$ V=\sum_{j=1}^4 b_jV^{(j)}(1)$ then the coefficients would stay fixed under the contraction. However, in terms of the original chosen basis the coefficients would transform as
$a_1=b_1,\ a_2=\frac{b_1}{\epsilon^4}-\frac{b_2}{2\epsilon ^3},\ a_3=b_3,\ a_4=-\frac{b_1}{\epsilon^2}+\frac{b_2}{2\epsilon}+b_4$.
\end{example}

\subsection{Contractions to the Heisenberg algebra}
For contractions to  nondegenerate or degenerate superintegrable systems formed from the Heisenberg algebra, our theorems concerning the potential do not apply, since the Heisenberg Hamiltonian is singular. In a paper to follow  we will  describe their forms. However, it is not difficult to work out each individual case and see that the induced contractions always exist.
\begin{example} 
$ S_9\to \ {\rm Heisenberg\ algebra}$: We use $\{ {\cal J}_1'+i{\cal J}_2',\  {\cal J}_1'-i{\cal J}_2',\  {\cal J}_3'\}=\{ \epsilon({\cal J}_1+i{\cal J}_2),\  {\cal J}_1-i{\cal J}_2,\ \epsilon {\cal J}_3\}$.
with  coordinate implementation $s_1=\frac{\cos\phi}{\cosh\psi}$, $s_2=\frac{\sin\phi}{\cosh\psi}$, $s_3=\frac{\sinh\psi}{\cosh\psi}$,
and substitutions  $\phi=i\epsilon\alpha-i\ln \sqrt{\epsilon}$, $\psi=\xi\sqrt{\epsilon}$,
 to\ get  ${\cal J}_3'=-ip_\alpha$, ${\cal J}_1'+i{\cal J}_2'=-i(\xi p_\alpha+p_\xi)$, ${\cal J}_1'-i{\cal J}_2'=i(-\xi p_\alpha+p_\xi)$. The contraction from $S_9$ is
${\cal H}'=-p_\alpha^2,\ -{\cal L}'_1=(\xi p_\alpha+p_\xi)^2+c_1\xi^2+\frac{a_3}{\xi^2}$,
$ {\cal L}_2'=p_\xi^2-\xi^2 p_\alpha^2+c_2\xi^2+\frac{a_3}{\xi^2}$,
where the potential parameters $a_j$ of $S_9$ contract as $a_1=-c_1/18\epsilon^4-c_2/8\epsilon^3$, $a_2=-c_1/18\epsilon^4+c_2/8\epsilon^3$. Note that there is no nonconstant potential $V$ but there are potential-like terms in the remaining symmetry generators. 
The contracted system is exactly the same as one obtains from the ansatz 
\[  {\cal H}'=-p_\alpha^2+V(\alpha,\xi),\ {\cal L}'_1=-(\xi p_\alpha+p_\xi)^2+W_1(\alpha,\xi), 
\ {\cal L}_2'=p_\xi^2-\xi^2 p_\alpha^2+W_2(\alpha,\xi),\]
by requiring  a nondegenerate quadratic algebra. The structure relation is
$ {\cal R'}^2=16\left({{\cal L}_1'}^2{\cal H}'-(c_2{\cal L}'_1+c_1{\cal L}_2')({\cal L}_1'+{\cal L}_2')-4c_1a_3{\cal H}'+a_3(c_2-c_1)\right)$.

\end{example}

\section{Conclusions and discussion}
The principal results obtained in this paper are as follows:
\begin{enumerate}
 \item We showed that there is a one-to-one correspondence between conjugacy classes of quadratic algebras in the enveloping algebras of $e(2,\C)$ and 
$o(3,\C)$, and isomorphism classes of 2nd order superintegrable systems with potential on constant curvature spaces. In effect, these Lie algebras ``know'' 
the classical and quantum superintegrable systems they can produce. Thus, the associated classical orbits and 
quantum special functions and their properties are derivable from the Lie algebras, even though the superintegrable systems may 
exhibit no group symmetry whatsoever. Part of the proof was based on a classification of all quadratic algebras up to conjugacy, and
we expect  to find a more compact, direct proof in the future.
\item We showed that Lie algebra contractions of $e(2,\C)$ and $o(3,\C)$, which are few in number and have long since been classified, 
 induce contractions of free quadratic algebras, and these in turn 
induce contractions of the corresponding classical and quantum superintegrable systems with potential. These algebraic contractions correspond to geometrical 
pointwise  limiting processes in the physical models. The procedure is rigid and deterministic. As shown in \cite{KMPost13}, one of the 
consequences of contracting between superintegrable systems is a series of limiting relations between special functions
associated with the superintegrable systems, a special case of which is the Askey scheme for hypergeometric orthogonal polynomials.
Again, part of the conclusions are based on step-by-step classification, which we expect to replace with a more compact proof.
\end{enumerate}

In  follow-up papers we will extend these results to all 2nd order 2D superintegrable systems, 
including those on Darboux and Koenig spaces. We shall also classify abstract quadratic algebras and their contractions,
including those not induced from Lie algebras, and study their relations with superintegrable systems.

\subsection*{Acknowledgment} Eyal Subag made important contributions to this paper through critical discussions.
This work was partially supported by a grant from the Simons Foundation (\# 208754 to Willard Miller, Jr.).

\bibliography{bib}{}

\begin{thebibliography}{10}

\bibitem{Conaster}
C.~W. Conatser.
\newblock Contractions of the low-dimensional real lie algebras.
\newblock {\em J. Math. Phys.}, 13:196--203, 1972.

\bibitem{Wigner}
E.~In\"on\"u and E.~P. Wigner.
\newblock On the contraction of groups and their representations.
\newblock {\em Proc. Nat. Acad. Sci.}, 39:510--524, 1953.

\bibitem{KMPost11}
E~G Kalnins, Miller.~W Jr, and Post S.
\newblock Two-variable wilson polynomials and the generic superintegrable
  system on the 3-sphere.
\newblock {\em SIGMA}, 7:051, 2011.

\bibitem{fine}
E.~G. Kalnins, J.~M. Kress, W.~{Miller Jr.}, and S.~Post.
\newblock Structure theory for second order 2d superintegrable systems with
  1-parameter potentials.
\newblock {\em SIGMA}, 46:085206, 2012.

\bibitem{KKMW}
E.~G. Kalnins, J.M. Kress, W.~{ Miller Jr. }, and P.~Winternitz.
\newblock Superintegrable systems in darboux spaces.
\newblock {\em J. Math. Phys.}, 44:5811--5848, 2003.

\bibitem{KKMP}
E.~G. Kalnins, J.M. Kress, W.~{Miller Jr. }, and G.~S. Pogosyan.
\newblock Completeness of superintegrability in two-dimensional constant
  curvature spaces.
\newblock {\em J. Math Phys.}, 34:4705--472, 2001.

\bibitem{KKM20041}
E.~G. Kalnins, J.M. Kress, and W.~{Miller Jr.}
\newblock Second order superintegrable systems in conformally flat spaces. i:
  2d classical structure theory.
\newblock {\em J. Math. Phys.}, 46:053509, 2005.

\bibitem{KKM20042}
E.~G. Kalnins, J.M. Kress, and W.~{Miller Jr.}
\newblock Second order superintegrable systems in conformally flat spaces. ii:
  The classical 2d {S}t\"ackel transform.
\newblock {\em J. Math. Phys.}, 46:053510, 2005.

\bibitem{KKM20061}
E.~G. Kalnins, J.M. Kress, and W.~{Miller Jr.}
\newblock Second order superintegrable systems in conformally flat spaces. v.
  2d and 3d quantum systems.
\newblock {\em J. Math Phys.}, 47:093501, 2006.

\bibitem{KMT}
E~G Kalnins, W~Jr Miller, and M~Tratnik.
\newblock Families of orthogonal and biorthogonal polynomials on the n-sphere.
\newblock {\em Families of orthogonal and biorthogonal polynomials on the
  n-sphere}, 22:272--294, 1991.

\bibitem{KMJP}
E.~G. Kalnins, W.~{Miller Jr. }, and G.~S. Pogosyan.
\newblock Superintegrability and associated polynomial solutions. {E}uclidean
  space and the sphere in two dimensions.
\newblock {\em J. Math. Phys}, 37:6439, 1996.

\bibitem{KMPost1}
E.~G. Kalnins, W.~{Miller Jr.}, and S.~Post.
\newblock Wilson polynomials and the generic superintegrable system on the
  2-sphere.
\newblock {\em J. Math Phys. A}, 40:11525--11538, 2007.

\bibitem{KMPost13}
E.~G. Kalnins, W.\ {Miller, Jr}, and S.\ Post.
\newblock Contractions of 2d 2nd order quantum superintegrable systems and the
  askey scheme for hypergeometric orthogonal polynomials.
\newblock {\em SIGMA}, 9:057, 28 pages, arXiv:1212.4766v1 [math--ph], 2013.

\bibitem{KMP}
E.G. Kalnins, Jr. W.~Miller, and G.S. Pogosyan.
\newblock Contractions of lie algebras and special function identities.
\newblock {\em J. Phys. A}, 32:4709--4732, 1999.

\bibitem{Koenigs}
G.~Koenigs.
\newblock {\em Lecons sur la th\'eorie g\'en\'erale des surfaces}, volume III,
  chapter Sur les g\'eod\'esiques a int\'egrales quadratiques, pages 368--404.
\newblock Chelsea Publishing, 1872.
\newblock In book by G. Darboux.

\bibitem{Kress2007}
J.~M. Kress.
\newblock Equivalence of superintegrable systems in two dimensions.
\newblock {\em Phys. Atomic Nuclei}, 70:560--566, 2007.

\bibitem{MPW}
W~Jr Miller, S~Post, and P~Winternitz.
\newblock Classical and quantum superintegrability with applications.
\newblock {\em J. Phys. A: Math. Theor.}, 46:42300, 2013.

\bibitem{NP}
M.~Nesterenko and R.~Popovych.
\newblock Contractions of low-dimensional lie algebras.
\newblock {\em J. Math. Phys.}, 47:123515--123515--45, 2006.

\bibitem{PW2011}
S.~Post and P.~Winternitz.
\newblock A nonseparable quantum superintegrable system in 2d real {E}uclidean
  space.
\newblock {\em J. Phys. A}, 44:152001, 2011.

\bibitem{Saletan}
E.. Saletan.
\newblock Contractions of lie groups.
\newblock {\em J. Math. Phys.}, 2:1--21, 1961.

\bibitem{TTW2}
F.~Tremblay, A.~V. Turbiner, and P.~Winternitz.
\newblock Periodic orbits for an infinite family of classical superintegrable
  systems.
\newblock {\em J. Phys. A: Math. Theor.}, 43:015202, 2010.

\bibitem{TTW}
F.~Tremblay, A.V. Turbiner, and P.~Winternitz.
\newblock An infinite family of solvable and integrable quantum systems on a
  plane.
\newblock {\em J. Phys. A: Math. Theor.}, 42:242001, 2009.

\bibitem{WW}
E.~Weimar-Woods.
\newblock The three-dimensional real lie algebras and their contractions.
\newblock {\em J. Math. Phys.}, 32:2028--2033, 1991.

\end{thebibliography}
\bibliographystyle{plain}

\end{document}